\newcount\draft\draft=0			
\newcount\cameraready\cameraready=1	
\pdfoutput=1
\documentclass[10pt,letterpaper]{sig-alternate-10pt}

\usepackage{graphicx,epsfig,sped,times,wide}
\usepackage{algorithm,url}
\usepackage{algpseudocode}
\newcommand\LComment[1]{\(\triangleright\) #1}%

\usepackage[noeka]{mathrmletter}
\usepackage{subfigure}
\usepackage{caption}
\usepackage{soul}
\usepackage{appendix}

\ifnum\draft=1
 \input{revision}
 \usepackage{drafthead}
\fi

\newcommand{\name} {Chronos}

\newenvironment{Itemize}%
{\begin{itemize}%
\setlength{\itemsep}{0pt}%
\setlength{\topsep}{0pt}%
\setlength{\partopsep}{0pt}%
\setlength{\parskip}{0pt}}%
{\end{itemize}}
{\begin{itemize}%
\setlength{\itemsep}{0pt}%
\setlength{\topsep}{0pt}%
\setlength{\partopsep}{0pt}%
\setlength{\parskip}{0pt}}%
{\end{itemize}}
\makeatletter
  \newcommand\figcaption{\def\@captype{figure}\caption}
  \newcommand\tabcaption{\def\@captype{table}\caption}
\makeatother

\newcommand{\xqed}{\nobreak \ifvmode \relax \else
      \ifdim\lastskip<1.5em \hskip-\lastskip
      \hskip1.5em plus0em minus0.5em \fi \nobreak
      \vrule height0.75em width0.5em depth0.25em\fi}

\newcommand{\xref}[1]{\S\ref{#1}}

\newcommand{\textred}[1]{\textcolor{red}{#1}}
\ifx\noeditingmarks\undefined
   \newcommand{\pgwrapper}[2]{\textred{#1: #2}}
\else
   \newcommand{\pgwrapper}[2]{}
\fi

\usepackage{xxx} 

\newcommand\sssection[1]{\vspace*{0.1cm}\noindent{\bf#1}}

\makeatletter

\global\def\section{\@startsection {section}{1}{\z@}%
                                   {2ex \@plus 1ex \@minus .1ex}%
                                   {1ex \@plus.2ex}%
                                   {\normalfont\bfseries\scshape\fontsize{11}{13}\selectfont}}
\global\def\subsection{\@startsection{subsection}{2}{\z@}%
                                     {2ex\@plus 1ex \@minus .1ex}%
                                     {1ex \@plus .2ex}%
                                     {\normalfont\bfseries\fontsize{10}{12}\selectfont}}
\global\def\subsubsection{\@startsection{subsubsection}{3}{\z@}%
                                     {2ex\@plus 1ex \@minus .1ex}%
                                     {1ex \@plus .2ex}%
                                     {\normalfont\itshape\fontsize{10}{12}\selectfont}}

\global\def\@maketitle{%
  \newpage
  \begin{center}%
  \let \footnote \thanks
  \null
    \vskip -.3em%
    {\bf\LARGE \@title \par}%
    \vskip 1em%
    {\large
      \lineskip .5em%
      \begin{tabular}[t]{c}%
        \@author
      \end{tabular}\par}%
    \vskip 1em%
    {\large \@date}%
  \end{center}%
  \par
  \vskip 2em}

\begin{document}

\title{ 
Sub-Nanosecond Time of Flight on Commercial Wi-Fi Cards
}

\newcommand{\supsym}[1]{\raisebox{6pt}{{\footnotesize #1}}}

\numberofauthors{1} 
\author{\alignauthor Deepak Vasisht, Swarun Kumar, and Dina Katabi \\
\affaddr{Massachusetts Institute of Technology}  \\
\email{\{deepakv,swarun,dk\}@mit.edu}
}

\date{}
\maketitle

\begin{sloppypar}

\begin{abstract}
Time-of-flight, i.e., the time incurred by a signal to travel from
transmitter to receiver, is perhaps the most intuitive way to measure
distances using wireless signals. It is used in major positioning
systems such as GPS, RADAR, and SONAR. However, attempts at using
time-of-flight for indoor localization have failed to deliver
acceptable accuracy due to fundamental limitations in measuring time
on Wi-Fi and other RF consumer technologies. While the research
community has developed alternatives for RF-based indoor localization
that do not require time-of-flight, those approaches have their own
limitations that hamper their use in practice. In particular, many
existing approaches need receivers with large antenna arrays while
commercial Wi-Fi nodes have two or three antennas. Other systems
require fingerprinting the environment to create signal maps. More
fundamentally, none of these methods support indoor positioning
between a pair of Wi-Fi devices without~third~party~support.

In this paper, we present a set of algorithms that measure the
time-of-flight to sub-nanosecond accuracy on commercial Wi-Fi
cards. We implement these algorithms and demonstrate a system
that achieves accurate device-to-device localization, i.e. enables a
pair of Wi-Fi devices to locate each other without any support from
the infrastructure, not even the location of the access points.

\end{abstract}
\section{Introduction}\label{sec:intro}
The time-of-flight of a signal captures the time it takes to
propagate from a transmitter to a receiver.  Time-of-flight is perhaps
the most intuitive method for localization using wireless signals. If
one can accurately measure the time-of-flight from a transmitter, one
can compute the transmitter's distance simply by multiplying the
time-of-flight by the speed of light. As early as World War I, SONAR
systems used the time-of-flight of acoustic signals to localize
submarines. Today, GPS, the most widely used outdoor localization
system, localizes a device using the time-of-flight of radio signals
from satellites.  However, applying the same concept to indoor
localization has proven difficult. Systems for localization in indoor
spaces are expected to deliver high accuracy (e.g., a meter or less)
using consumer-oriented technologies (e.g., Wi-Fi on one's cellphone).
Unfortunately, past work could not measure time-of-flight at such an
accuracy on Wi-Fi devices~\cite{tof1,tof2}.  As a result, over the
years, research on accurate indoor positioning has moved towards more
complex alternatives such as employing large multi-antenna arrays to
compute the angle-of-arrival of the signal~\cite{arraytrack,
  PinPoint}. These new techniques have delivered highly accurate
indoor localization systems.

\begin{figure}
\centering		
\includegraphics[width=0.45\textwidth, height=0.2\textheight]{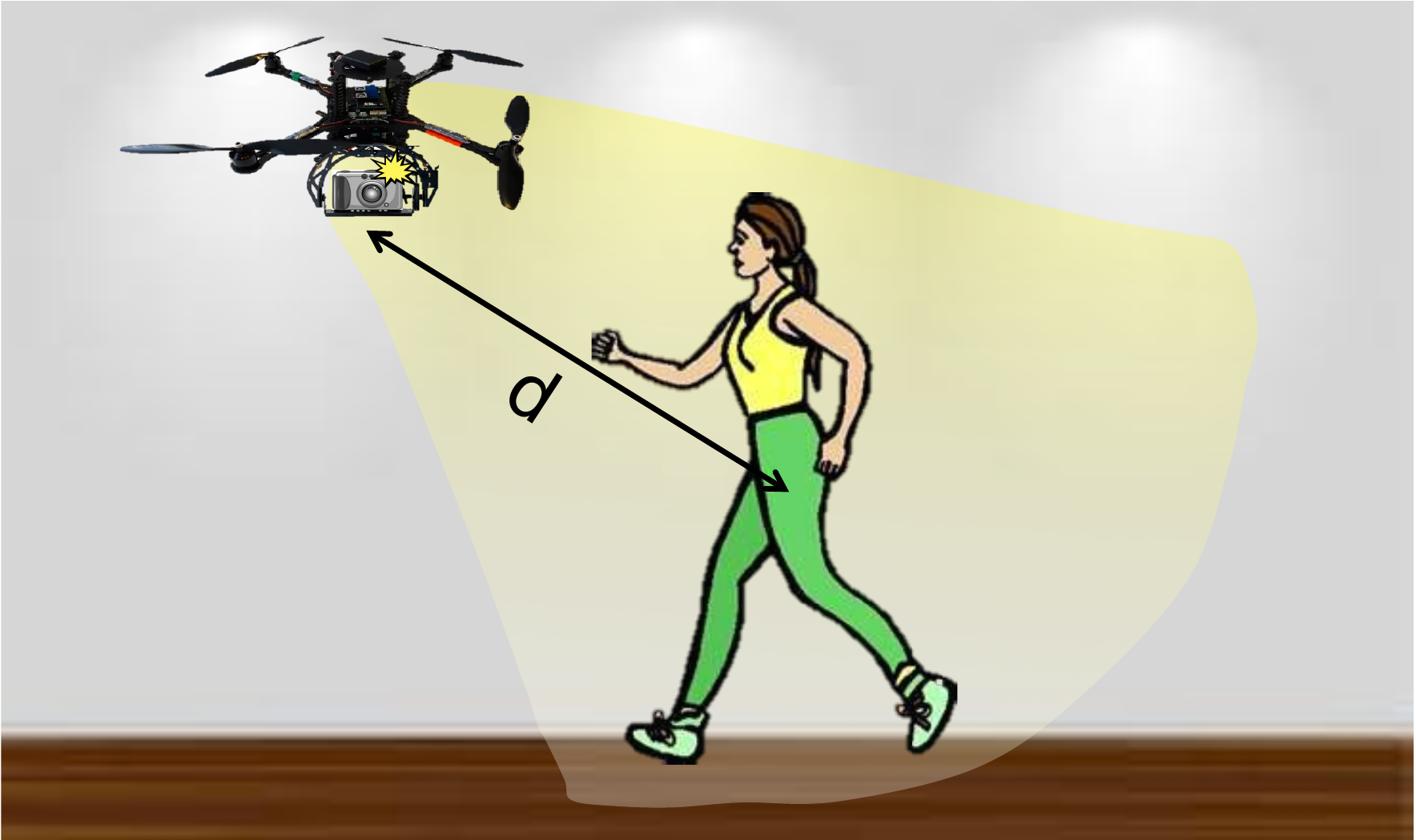}
\vspace*{-0.05in}
\caption{\footnotesize {\bf \name\ on a personal drone:}
  \name\ enables a drone to maintain a safe distance $d$ from a user
  by tracking a device in her pocket, while capturing a video in
  uncalibrated indoor environments. }
\label{fig:intropic}
\vspace*{-0.2in}
\end{figure}

Despite these advances, time-of-flight based localization has some of the basic
desirable features that state-of-the-art indoor localization systems
lack.  In particular, measuring time-of-flight does not require more
than a single antenna on the receiver.  In fact, by measuring
time-of-flight of a signal to just two antennas, a receiver can
intersect the corresponding distances to locate its source.  In other
words, a receiver can locate a wireless transmitter with no support
from surrounding infrastructure whatsoever. This is quite unlike
current indoor localization systems, which require the help of
multiple access points to find the distance between a pair of mobile
devices. Furthermore, each of these access points need to have many
antennas -- far beyond what is supported in commercial Wi-Fi
devices. In addition, the location of these access points has to be
calibrated and known a priori.

But, why is it that one cannot accurately measure time-of-flight on
commercial Wi-Fi devices in the first place? In particular, to achieve
state-of-the-art positioning accuracy, one must measure
time-of-flight at sub-nanosecond granularity. However, doing so on
commercial Wi-Fi cards is fundamentally challenging for the following
three reasons.

\sssection{Limited Time Granularity: } First, the straightforward
approach to measure time-of-flight is to read off the clock of the
Wi-Fi radio when the signal arrives~\cite{tof1}. Unfortunately, the
clocks on today's Wi-Fi cards operate at tens of Megahertz, limiting
their resolution in measuring time to tens of nanoseconds~\cite{tof2,
  tof3, tof5}. To put this in perspective, a clock running at 20~MHz
(the bandwidth of typical Wi-Fi systems), can only tell apart
distances separated by 15~m, making it impractical for accurate indoor
positioning. Even recent state-of-the-art systems that measure
time-of-flight using high-resolution 88~MHz Wi-Fi clocks~\cite{sail}
and super-resolution channel processing
techniques~\cite{synchronicity} suffer a mean localization error of
about 2.3~m.

\sssection{Packet Detection Delay: } Second, any measurement of
time-of-flight of a packet necessarily includes the delay in detecting
its presence. To make matters worse,~this packet detection delay is
typically orders-of-magnitude higher than time-of-flight. For indoor
Wi-Fi environments, time-of-flight is just a few nanoseconds, while
packet detection delay spans hundreds of
nanoseconds~\cite{sourcesync}.  Today, there is no way to tease apart
the time-of-flight from this detection delay.

\sssection{Multipath: } Finally, in indoor environments, signals do
not experience a single time-of-flight, but a time-of-flight
spread. To see why, observe that wireless signals in indoor
environments travel along multiple paths, and bounce off walls and
furniture. As a result, the receiver obtains several copies of the
signal, each having experienced a different time-of-flight. To perform
accurate localization, one must therefore be able to disentangle the
time-of-flight of the most direct path from all the remaining paths.

{In this paper, we show that it is possible to design algorithms that
  overcome the above limitations and measure the time-of-flight at
  sub-nanosecond accuracy using off-the-shelf Wi-Fi cards. At a high
  level, our approach is based on the following observation: If one
  had a very wideband radio (e.g., a few GHz), one could measure time
  of flight at sub-nanosecond accuracy. While each Wi-Fi frequency
  band is only tens of Megahertz wide, there are many such bands that
  together span a very wide bandwidth.  Our solution therefore
  collects measurements on multiple Wi-Fi frequency bands and stitches
  them together to give the illusion of a wide-band radio. Our key
  contribution is an algorithm that achieves this, despite the fact
  that Wi-Fi frequency bands are non-contiguous, and in some cases, a
  few Gigahertz apart. We further develop a set of algorithms that
  build on this idea to overcome each of the aforementioned
  challenges.  We also detail the benefits and limitations of such a
  design.}


To demonstrate the performance and practicality of our design, we
built \name, a software-only solution that harnesses our algorithms to
enable a pair of commercial Wi-Fi devices to locate each other without
any support from the infrastructure. To illustrate its capabilities,
we apply \name\ to personal drones~\cite{personal} that follow a user
around and capture videos of their everyday indoors activities. Such
drones can help monitor fitness, activities and exercise of users at
home, work or the gym (see Fig.~\ref{fig:intropic}). \name\ allows a
personal drone to maintain the best possible distance relative to its
user to take optimal videos at the right focus. It achieves this by
using the Wi-Fi card on the drone to locate the user's device, without
any help from the infrastructure. {The application also illustrates
  \name's ability to run on standard 3-antenna Wi-Fi cards, as opposed
  to large antenna arrays, which would be too heavy and difficult to
  mount on lightweight indoor drones.}

We evaluated \name's performance on pairs of devices equipped with
Intel 5300 Wi-Fi cards, including Thinkpad W530 laptops, as well as an
AscTec Atom board (a small computing board) mounted on an AscTec
Hummingbird Quadrotor drone platform.
Our results reveal the following:
\begin{Itemize}
\item \name\ achieves a median error in time-of-flight of 0.47 ns in
  line-of-sight and 0.69 ns in non-line-of-sight settings,
  corresponding to a physical distance accuracy of 14.1 ~cm and 20.7
  ~cm respectively.
\item \name\ uses time-of-flight to triangulate the location of the
  device with a median error of 58 cm in line-of-sight and 118~cm in
  non-line-of-sight settings.
\item When mounted on a drone, \name\ integrates with robotic control
  algorithms to further improve its accuracy. It maintains the
  required distance relative to a user's device with a root
  mean-squared error of 4.2 cm.
\end{Itemize}

\noindent \textbf{Contributions: } To our knowledge, \name\ is the
first RF-based positioning system that can measure sub-nanosecond time
of flight on commercial Wi-Fi cards. \name\ leverages the
time-of-flight measurements to estimate device-to-device distance
measurements without any infrastructure support. Finally,
\name\ operates on typical 2/3-antenna Wi-Fi receivers, yet delivers
state-of-the-art~localization~accuracy.



\section{Related Work}\label{sec:related_work}
This paper is closely related to past work that measures the
time-of-flight of Wi-Fi signals. There have been several studies that
resolve time-of-flight to around ten nanoseconds using the clocks of
Wi-Fi cards~\cite{tof1, tof4, caesar, conexttof}.  Many conclude that
the clocks on current Wi-Fi hardware alone cannot permit higher
resolutions of time-of-flight~\cite{tof2, tof3, tof5}. Some systems
have attempted to compensate for the lack of accurate time-of-flight
measurements on Wi-Fi radios by augmenting their designs with other
sensors and hardware. In particular, SAIL~\cite{sail} couples
time-of-arrival measurements on the 88~MHz clock of an Atheros Wi-Fi
card with inertial motion sensors on a mobile device. It asks the user
to physically walk to different locations and couples Wi-Fi channel
measurements at a single access point with readings of motion sensors
on their mobile device. SAIL processes this information to measure
time-of-flight at a granularity of several nanoseconds, achieving
localization accuracy of a few meters. However, unlike \name, SAIL
requires users to physically move to different locations, along
restricted classes of trajectories, due to its reliance on motion
sensors. Synchronicity~\cite{synchronicity} uses three WARP access
points to compute the location of a Wi-Fi transmitter using their
time-difference of arrival. Synchronicity requires the different
access points to be synchronized in time. The authors achieve this in
their current implementation by connecting the access points to the
same reference clock, and leave distributed time-synchronization for
future work. We believe \name\ coupled with
SourceSync~\cite{sourcesync} can complement Synchronicity by
maintaining accurate time-synchronization between access points, while
accounting for their relative time-of-flight. Finally recent
theoretical work~\cite{sap1,theorytoa} has proposed using a single
large 8-antenna array to measure time-of-flight for indoor
positioning. Unlike \name, these papers assume time and frequency
synchronization of the access point and client, which is hard to
ensure in practice~\cite{megamimo}.

Our work is also related to other RF-based indoor localization
solutions. Such systems measure metrics other than time-of-flight,
like angle-of-arrival and received signal power, across many RF
receivers in the environment. Some achieve this using advanced
infrastructure such as antenna arrays~\cite{PinPoint, arraytrack,
  aoaest}. Others rely on a combination of fingerprinting of the
environment and modeling received signal power at multiple client
locations using multiple access points in the environment~\cite{zee,
  ez, radar, horus}. Recent work requires neither~\cite{ubicarse,
  phaser}, but assumes the presence of multiple Wi-Fi access points in
the environment. Unlike these systems, \name\ infers location between
a single pair of commodity Wi-Fi devices, without requiring prior
fingerprinting of the environment or support of the infrastructure.

This paper is related to past systems on time-synchronization. For
example, SourceSync~\cite{sourcesync} and FICA~\cite{fica} measure
time-of-arrival to synchronize the transmissions of distributed access
points. However, these systems mainly focus on estimating
time-of-arrival as opposed to time-of-flight, which is dominated by
packet detection delay, as we show in~\xref{sec:pdelay}. In contrast,
\name\ directly measures time-of-flight at a sub-nanosecond
granularity, bereft of packet detection delay, and can therefore
complement these systems to further improve their accuracy.

Finally, our work relates to past non-Wi-Fi localization systems, some
measuring time-of-flight, e.g. ultra-wideband~\cite{harmonica}, pulse
radar~\cite{pulse}, acoustic systems~\cite{cricket, guoguo},
device-free localization systems~\cite{witrack, wisee,
  singledevicefree}, vision-based systems~\cite{tracksense, footslam,
  wifislam} and inertial-measurement based systems~\cite{lietal,
  adapt}. These systems either deploy custom
infrastructure~\cite{cricket,pulse}, assume special markers in the
environment~\cite{markers}, or suffer from poor
accuracy~\cite{adapt}. In contrast, \name\ can leverage existing Wi-Fi
radios that are ubiquitous in today's mobile devices, laptops and
access points.

\section{Overview of \name}\label{sec:overview}
This section briefly outlines \name's key challenges. \name's core
contribution is a new method that computes time of flight of a Wi-Fi
signal.  However, as mentioned in the introduction, obtaining highly
accurate time-of-flight on Wi-Fi devices requires addressing three
main challenges:
\begin{Itemize}
\item \textbf{Limited Time Granularity: } First, \name\ needs to
  compute accurate time-of-flight despite the limited clock
  resolution of commercial Wi-Fi cards (See~\xref{sec:tof}).
\item \textbf{Eliminating Packet Detection Delay: } Second, it must
  disentangle time-of-flight from packet detection delay, which is
  often orders-of-magnitude larger (See~\xref{sec:pdelay}).
\item \textbf{Combating Multipath: } Third, \name\ should separate the
  time-of-flight of direct path of the wireless signal from that of
  all the remaining paths (See~\xref{sec:multipath}).
\end{Itemize}
In the following sections, we describe how \name\ overcomes each of
the above challenges as well as other practical issues to enable a
robust system design.

\begin{figure}
\centering		
\includegraphics[width=0.45\textwidth]{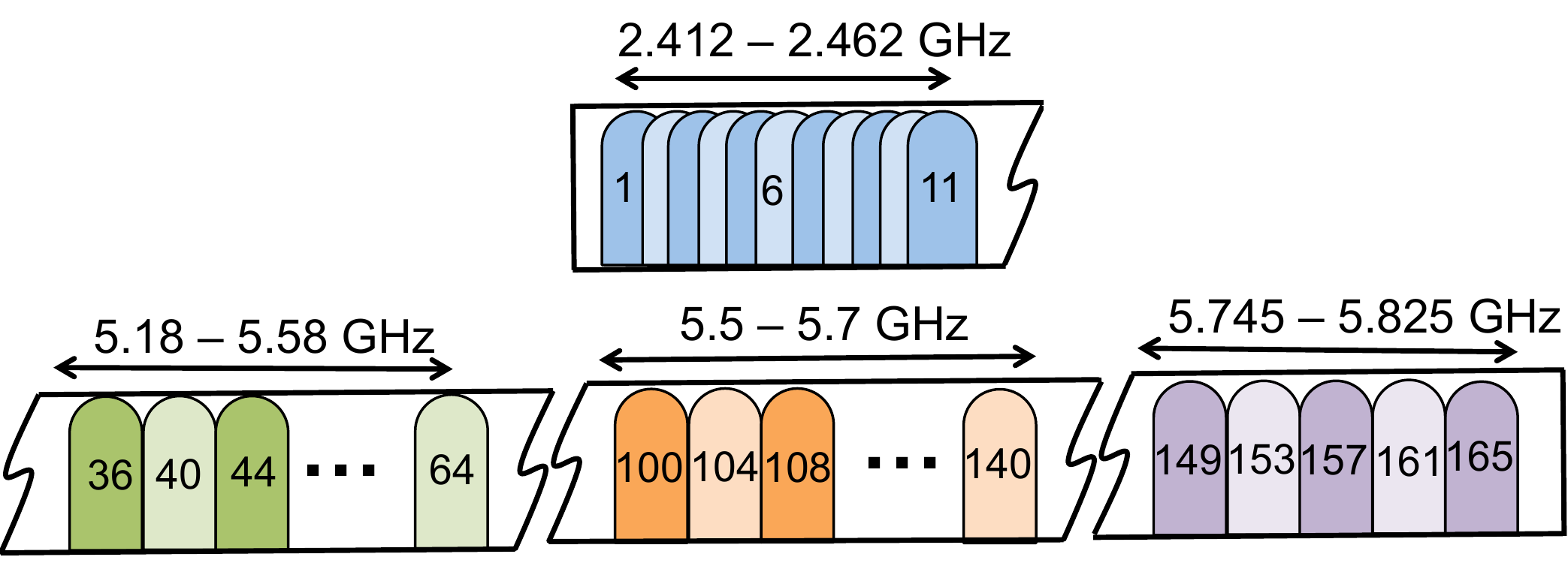}
\vspace*{-0.05in}
\caption{\footnotesize{\bf Wi-Fi Bands:} Depicts Wi-Fi bands at 2.4~GHz and
  5~GHz. Note that some of these frequencies (e.g.~5.5-5.7~GHz) are
  DFS bands in the U.S. that many 802.11h compatible 802.11n radios
  like Intel 5300 support. }
\label{fig:wifibands}
\vspace*{-0.15in}
\end{figure}

\section{Measuring Time of Flight}\label{sec:tof}
In this section, we describe how \name\ measures accurate
time-of-flight of received signals, despite the limited time
resolution of commodity Wi-Fi devices. For clarity, the rest of this
section assumes signals propagate from the transmitter to a receiver
along a single path with no detection delay. We address challenges
stemming from packet detection delay and multipath
in~\xref{sec:pdelay} and~\xref{sec:multipath} respectively.

\name's approach is based on the following observation: Conceptually,
if our receiver had a very wide bandwidth, it could readily measure
time-of-flight at a fine-grained resolution. Unfortunately, today's
Wi-Fi devices do not have such wide bandwidth. But there is another
opportunity: Wi-Fi devices are known to span multiple frequency bands
scattered around 2.4~GHz and 5~GHz. {Combined, these bands
  span almost one GHz of bandwidth.} By making a transmitter and
receiver hop between these different frequency bands, we can gather
many different measurements of the wireless channel. We can then
``stitch together'' these measurements to compute the time-of-flight,
as if we had a very wideband radio.

{While channel hopping provides the intuition on how to compute
  accurate time-of-flight, transforming this idea into practice is not
  simple. Our method must account for the fact that many Wi-Fi bands
  are non-contiguous, unequally spaced, and even multiple GHz apart
  (see Fig.~\ref{fig:wifibands}).}

\name's solution to overcome these challenges exploits the {\it phase}
of wireless channels. Specifically, we know from basic
electromagnetics that as a signal propagates in time, it accumulates a
corresponding phase depending on its frequency. The higher the
frequency of the signal, the faster the phase accumulates. To
illustrate, let us consider a transmitter sending a signal to its
receiver. Then we can write the wireless channel $h$ as~\cite{Tse}:
\begin{align}
h = a e^{-j 2\pi f \tau} \label{eqn:basicchan}
\end{align}
Where $a$ is the signal magnitude that captures its attenuation over
the air, $f$ is the frequency and $\tau$ is the time-of-flight. The
phase of this channel depends on time-of-flight as:
\begin{align}
\angle h = - 2\pi f \tau \mod 2\pi \label{eqn:channel}
\end{align}
Notice that the above equation depends directly on the signal's
time-of-flight. In other words, it does not depend on the signal's
precise time-of-departure at the transmitter. Hence, we can use
Eqn.~\ref{eqn:channel} above to measure the time-of-flight $\tau$ as:
\begin{align}
\tau =  - \frac{\angle h}{2 \pi f} \mod \frac{1}{f}
\end{align}

The above equation gives us the time-of-flight modulo
${1}/{f}$. Hence, for a Wi-Fi frequency of $2.4$~GHz, we can only
obtain the time-of-flight {\it modulo} $0.4$~nanoseconds. Said
differently, transmitters with times-of-flight $0.1$~ns, $0.5$~ns,
$0.9$~ns, $1.3$~ns, etc. all produce identical phase in the wireless
channel. In terms of physical distances, this means transmitters at
distances separated by multiples of $12$~cm (e.g., $3$~cm, $15$~cm,
$27$~cm, $39$~cm, etc.)  all result in the same channel
phase. Consequently, there is no way to distinguish between these
transmitters using their phase on a single channel.

Indeed, this is precisely why \name\ needs to hop between multiple
frequency bands $\{f_1, \dots, f_n\}$ and measure the corresponding
wireless channels $\{h_1, \dots, h_n\}$. The result is a system of
equations, one per frequency, that measure the time-of-flight modulo
different values:
\begin{align}
\tau =  & - \frac{\angle h_1}{2 \pi f_1} \mod \frac{1}{f_1} \nonumber\\
\tau =  & - \frac{\angle h_2}{2 \pi f_2} \mod \frac{1}{f_2} \nonumber\\
& \vdots\nonumber\\
\tau =  & - \frac{\angle h_{n}}{2 \pi f_n} \mod \frac{1}{f_{n}} \label{eqn:chinese}
\end{align}

Notice that the above set of equations has the form of the well-known
Chinese remainder theorem~\cite{chinesefourier}. Such equations can be
readily solved using standard modular arithmetic algorithms, even
amidst noise~\cite{chinese}.\footnote{Algorithm~\ref{alg:ndft}
  in~\xref{sec:multipath} provides a more general version of \name's
  algorithm to do this while accounting for noise and multipath} The
theorem states that solutions to these equations are unique modulo a
much larger quantity -- the Least Common Multiple (LCM) of
$\{{1}/{f_1}, \dots, {1}/{f_n}\}$. For instance, \name\ can resolve
time-of-flight uniquely modulo 200~ns using Wi-Fi frequency bands
around 2.4~GHz.  That is, it can resolve transmitters closer than
$60~\text{m}$ in distance without ambiguity, which is sufficient for
most indoor environments.

\begin{figure}
\centering		
\includegraphics[width=0.46\textwidth,height=0.23\textheight]{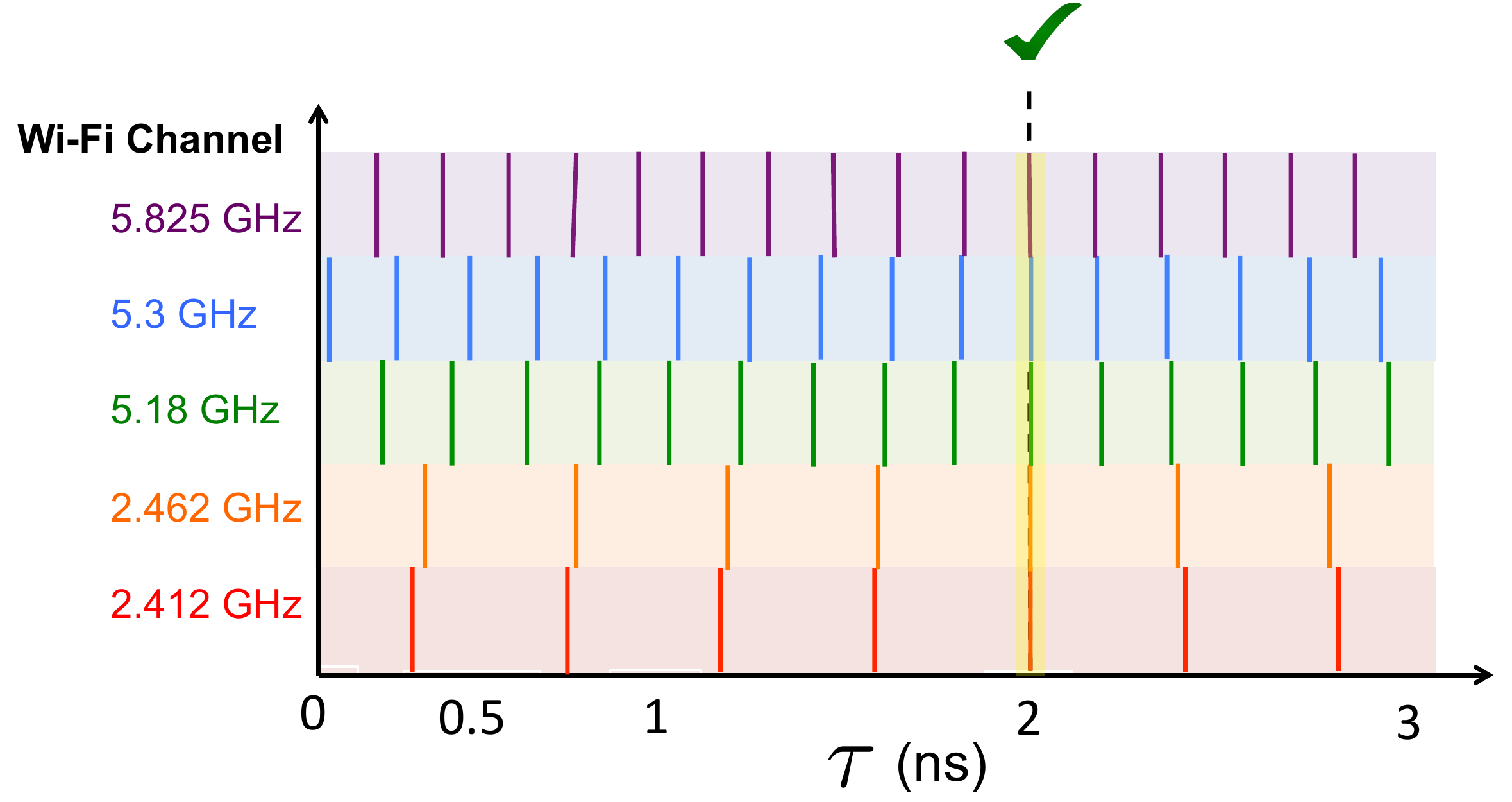}
\vspace*{-0.05in}
\caption{\footnotesize {\bf Measuring Time-of-Flight:} Consider a wireless
  transmitter at a distance of $0.6$ m, i.e. a time-of-flight of
  $2$~ns. The phase of each Wi-Fi channel results in multiple
  solutions, depicted as colored lines, including $2$~ns. However, the
  solution that satisfies most equations, i.e. has the most number of
  aligned colored lines is the true time-of-flight ($2$~ns). }
\label{fig:chinese}
\vspace*{-0.15in}
\end{figure}

To illustrate how the above system of equations works, consider a
source at 0.6~m whose time-of-flight is 2~ns. Say the receiver
measures the channel phases from this source on five candidate Wi-Fi
channels as shown in Fig.~\ref{fig:chinese}. We note that a
measurement on each of these channels produces a unique equation for
$\tau$, like in Eqn.~\ref{eqn:chinese}. Each equation has multiple
solutions, depicted as colored vertical lines in
Fig.~\ref{fig:chinese}. However, only the correct solution of $\tau$
will satisfy all equations. Hence, by picking the solution satisfying
the most number of equations (i.e., the $\tau$ with most number of
aligned lines in Fig.~\ref{fig:chinese}), we can recover the true
time-of-flight of 2~ns.

Note that our solution based on the Chinese remainder theorem makes no
assumptions on whether the set of frequencies $\{f_1, \dots, f_n\}$
are equally separated or otherwise. In fact, having unequally
separated frequencies makes them less likely to share common factors,
boosting the LCM. This means that counter-intuitively, the scattered
and unequally-separated bands of Wi-Fi (see Fig.~\ref{fig:wifibands})
are not a challenge, but an opportunity to resolve larger values of
$\tau$.

While the above provides a mathematical formulation of our algorithm,
we describe below important systems considerations in applying
\name~to~commercial~Wi-Fi~cards:
\begin{Itemize}  
\item \name\ must ensure both the Wi-Fi transmitter and receiver hop
  synchronously between multiple Wi-Fi channels.  \name\ achieves this
  using a channel hopping protocol driven by the transmitter. Before
  switching frequency bands (every 2-3~ms in our implementation), the
  transmitter issues a control packet that advertises the frequency of
  the next band to hop to. The receiver responds with an
  acknowledgment and switches to the advertised channel. Once the
  acknowledgment is received, the transmitter switches frequency bands
  as well. As a fail-safe, transmitters and receivers revert to a
  default frequency band if they do not receive packets or
  acknowledgments from each other for a given time-out duration on any
  band. 
\item Our implementation of \name\ sweeps all Wi-Fi bands in 84~ms (12
  times per second). This is within the channel coherence time of
  indoor environments~\cite{megamimo} and can empirically localize
  users at walking speeds (see~\xref{sec:droneresults}).
\item \name\ primarily targets device-to-device localization between
  two users where data is typically not exchanged. However, we discuss
  and evaluate the implications of \name's protocol on data traffic
  in~\xref{sec:limitations} and~\xref{sec:tcpeffect}.
\item Finally, wireless transmitters and receivers experience carrier
  frequency offsets (CFO). These offsets cause phase errors in
  measured wireless channels.~\xref{sec:practical} describes how
  \name\ corrects frequency offsets, and additional phase offsets from
  differences in transmit and receive hardware.
\end{Itemize}

\section{Eliminating Packet Detection Delay} \label{sec:pdelay}
Our discussion so far has computed time-of-flight based on the
wireless channels $h_i$ that signals experience when transmitted over
the air on different frequencies $f_i$. The phase of such channels
depends exclusively on the time-of-flight of the signal, and its
frequency. In practice however, the \textit{measured} wireless
channels at the receiver, $\tilde{h}_i$, experience a delay in
addition to time-of-flight: the delay in detecting the presence of a
packet. This delay occurs because Wi-Fi receivers detect the presence
of a packet based on the energy of its first few time samples. The
number of samples that the Wi-Fi receiver needs to cross its energy
detection threshold varies based on the power of the received signal,
as well as noise. While this variation may seem small, packet
detection delays are often an order-of-magnitude larger than
time-of-flight, particularly in indoor environments, where
time-of-flight is just a few tens of nanoseconds
(See~\xref{sec:restof}).

Hence, our main goal here is to derive the true channel $h_i$
{(which incorporates the time-of-flight alone) from the
  measured channel $\tilde{h}_i$ (which incorporates both
  time-of-flight and packet detection delay)}. To do this, we exploit
the fact that Wi-Fi uses OFDM. Specifically, the bits of Wi-Fi packets
are transmitted in the frequency domain on several small frequency
bins called OFDM subcarriers. This means that the wireless channels
$\tilde{h}_i$ can be measured on each subcarrier. We then make the
following main claim: The measured channel at subcarrier-$0$ does not
experience packet detection delay, i.e. it is identical in phase to
the true channel at subcarrier $0$.

To see why this claim holds, note that while time-of-flight and packet
detection delay appear very similar, they occur at different stages of
a signal's lifetime. Specifically, time-of-flight occurs while the
wireless signal is transmitted over the air (i.e., in passband). In
contrast, packet detection delay stems from energy detection that
occurs in digital processing once the carrier frequency has been
removed (in baseband). Thus, time-of-flight and packet detection delay
affect the wireless OFDM channels in slightly different ways.

To understand this difference, consider a particular Wi-Fi frequency
band $i$. Let $\tilde{h}_{i, k}$ be the measured channel of OFDM
subcarrier $k$, at frequency $f_{i, k}$.  $\tilde{h}_{i, k}$
experiences two phase rotations in different stages of the signal's
lifetime:
\begin{Itemize}
\item A phase rotation in the air proportional to the over-the-air
  frequency $f_{i, k}$, just like the true wireless channel $h_{i,
    k}$. From Eqn.~\ref{eqn:channel} in~\xref{sec:tof}, this phase
  value for a frequency $f_{i, k}$ is:
     $$\angle h_{i, k} = -2\pi f_{i, k} \tau,$$
where $\tau$ is the time-of-flight.
\item An additional phase rotation due to packet detection after the
  removal of the carrier frequency. This additional phase rotation can
  be expressed in a similar form as:
   $$\Delta_{i,k} =  - 2\pi (f_{i, k} - f_{i, 0}) \delta_i,$$  
where $\delta_i$ is the packet detection delay. 
\end{Itemize}
Thus, the total measured channel phase at subcarrier $k$ is:
\begin{align}
\angle \tilde{h}_{i, k} = &  \angle h_{i,k} + \Delta_{i, k}\\
                        = & {-2\pi} f_{i, k} \tau - 2 \pi (f_{i, k} - f_{i, 0}) \delta_i  \label{eqn:hanglefin}
\end{align}
Notice from the above equation that the second term $\Delta_{i, k} = -
2\pi (f_{i, k} - f_{i, 0}) \delta_i = 0$ at precisely $k = 0$. In
other words, at the zero-subcarrier of OFDM, the measured channel
$\tilde{h}_{i,k}$ is identical in phase to the true channel $h_{i, k}$
over-the-air which validates our claim. 

In practice, this means that we can apply the Chinese Remainder
theorem as described in Eqn.~\ref{eqn:chinese} of~\xref{sec:tof}
at the zero-subcarriers (i.e. center frequencies) of each Wi-Fi
frequency band. In the U.S., Wi-Fi at 2.4~GHz and 5~GHz has a total of
35 Wi-Fi bands with independent center frequencies.\footnote{Including
  the DFS bands at 5~GHz in the U.S. which are supported by many
  802.11h-compatible 802.11n radios like the Intel 5300.} Therefore, a
sweep of all Wi-Fi frequency bands results in 35 independent equations
like in Eqn.~\ref{eqn:chinese}, which we can solve to recover
time-of-flight.

However, one problem still needs to be addressed. So far we have used
the measured channel at the zero-subcarrier of Wi-Fi bands. However,
Wi-Fi transmitters do not send data on the zero-subcarrier, meaning
that this channel simply cannot be measured.  This is because the
zero-subcarrier overlaps with DC offsets in hardware that are
extremely difficult to remove~\cite{ofdmbook, 80211standard}. So how
can one measure channels on zero-subcarriers if they do not even
contain data?

Fortunately, \name\ can tackle this challenge by using the remaining
Wi-Fi OFDM subcarriers, where signals are transmitted. Specifically,
it leverages the fact that indoor wireless channels are based on
physical phenomena. Hence, they are continuous over a small number of
OFDM subcarriers~\cite{chancont}. This means that \name\ can
interpolate the measured channel phase across all subcarriers to
estimate the missing phase at the zero-subcarrier.\footnote{Our
  implementation of \name\ uses cubic spline interpolation.} Indeed,
the 802.11n standard~\cite{80211standard} measures wireless channels
on as many as 30 subcarriers in each Wi-Fi band. Hence, interpolating
between the channels not only helps \name\ retrieve the measured
channel on the zero-subcarrier, but also provides additional
resilience to noise.

To summarize, \name\ applies the following steps to account for packet
detection delay: (1) It obtains the measured wireless channels on the
30 subcarriers on the 35 available Wi-Fi bands; (2) It interpolates
between these subcarriers to obtain the measured channel phase on the
zero-subcarriers on each of these bands, which is unaffected by packet
detection delay. (3) It retrieves the time-of-flight using the
resulting 35 channels.



\section{Combating Multipath} \label{sec:multipath}
\begin{figure}
\centering		
\includegraphics[width=0.45\textwidth]{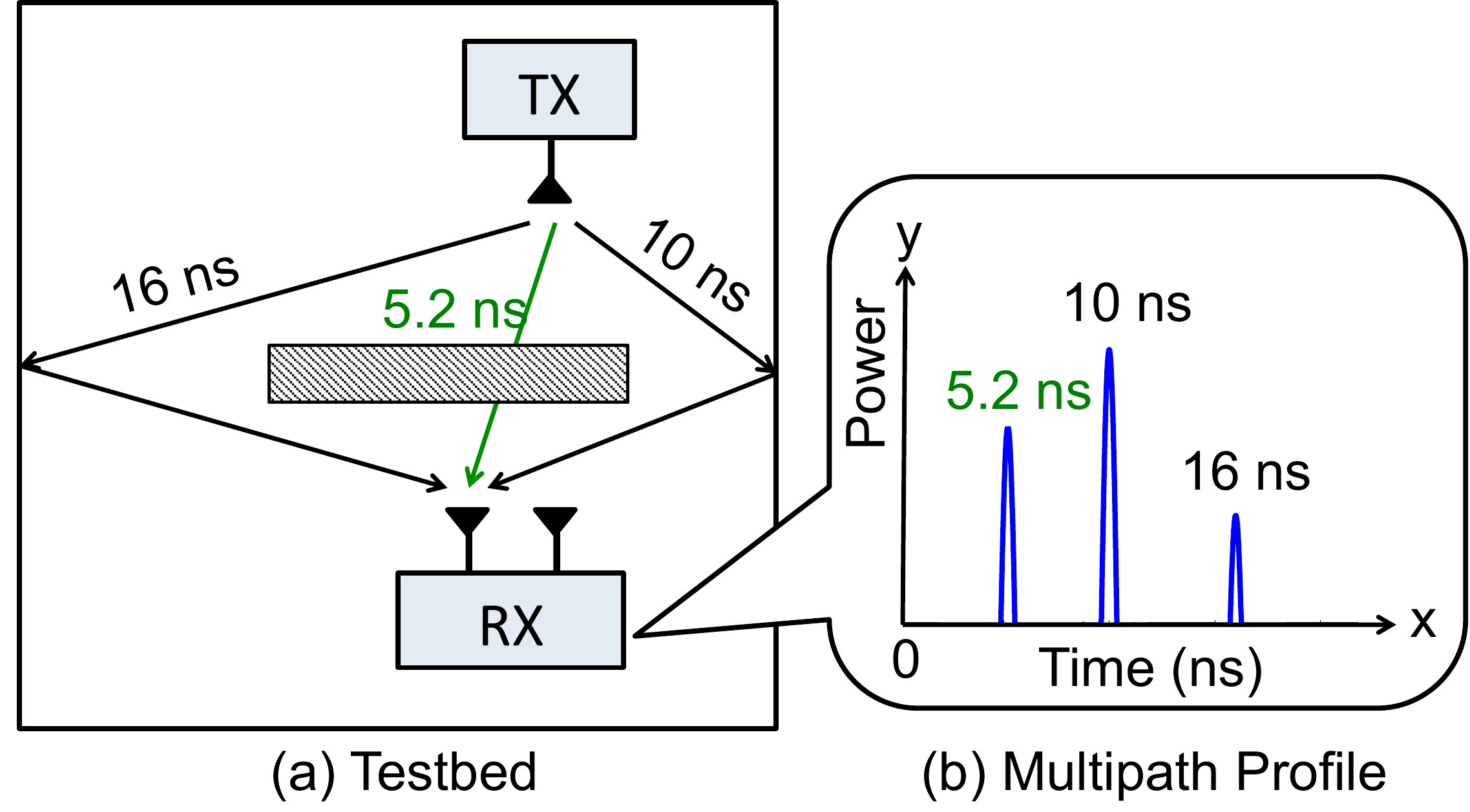}
\caption{\footnotesize{\bf Combating Multipath:} Consider a signal propagating from
  a transmitter to a receiver along three paths as shown in (a): an
  attenuated direct path and two reflected paths of lengths 5.2~ns,
  10~ns and 16~ns respectively. These paths can be separated by using
  the inverse discrete Fourier transform as shown in (b). The plot has
  three peaks corresponding to the propagation delays of the three
  paths, with peak magnitudes scaled by their relative attenuations. }
\label{fig:multipath}
\vspace*{-0.1in}
\end{figure}

So far, our discussion has assumed that a wireless signal propagates
along a single direct path between its transmitter and
receiver. However, indoor environments are rich in multipath, causing
wireless signals to bounce off objects in the environment like walls
and furniture. Fig.~\ref{fig:multipath}(a) illustrates an example
where the signal travels along three paths from its sender to
receiver. The signals on each of these paths propagate over the air
incurring different time delays as well as different attenuations. The
ultimate received signal is therefore the sum of these multiple signal
copies, each having experienced a different propagation
delay. Fig.~\ref{fig:multipath}(b) represents this using a \textit{multipath
profile}. This profile has peaks at the propagation delays of signal
paths, scaled by their respective attenuations. Hence, \name\ needs a
mechanism to find such a multipath profile, so as to separate the
propagation delays of different signal paths. This allows it to then
identify the time-of-flight as the least of these propagation delays,
i.e. the delay of the most direct (shortest) path.


\subsection{Computing Multipath Profiles}
Let us assume that wireless signals from a transmitter reach a
receiver along $p$ different paths. The received signal from each path
corresponds to amplitudes $\{a_1, \dots, a_p\}$ and propagation delays
$\{\tau_1, \dots, \tau_p\}$.  Observe that our earlier
Eqn.~\ref{eqn:basicchan}, considers only a single path experiencing
propagation delay and attenuation. In the presence of multipath, we
can extend this equation to write the measured channel $\tilde{h}_{i,
  0}$ on center-frequency $f_{i,0}$ as the sum of the channels on each
of these paths, i.e.:
\begin{align}
\tilde{h}_{i, 0} = \sum_{k=1}^p a_k e^{-j 2\pi f_{i, 0} \tau_k}  \text{~~~~~, for $i = 1, \dots, n$}\label{eq:h}
\end{align}

At this point, we need to disentangle these different paths and
recover their propagation delays. To do this, notice that the above
equation has a familiar form -- it is the well-known Discrete Fourier
Transform. Thus, if one could obtain the channel measurements at many
uniformly-spaced frequencies, a simple inverse-Fourier transform would
separate individual paths.  Such an inverse Fourier transform has a
closed-form expression that can be used to obtain the propagation
delay of all paths and compute the multipath profile (up to a
resolution defined by the bandwidth).
 
Wi-Fi frequency bands, however, are not equally spaced -- they are
scattered around 2.4~GHz and multiple non-contiguous chunks at 5~GHz,
as shown in Fig.~\ref{fig:wifibands}. While we can measure
$\tilde{h}_{i, 0}$ at each Wi-Fi band, these measurements will not be
at equally spaced frequencies and hence cannot be simply used to
compute the inverse Fourier transform.  In fact, since our
measurements of the channels are not uniformly spaced, we are dealing
with the \textit{Non-uniform} Discrete Fourier Transform or
NDFT~\cite{ndft}. To recover the multipath profile, we need to invert
the NDFT.

\subsection{Inverting the NDFT} \label{sec:ndft}

{To find the multipath profile, we must invert a Non Uniform Discrete
  Fourier Transform (NDFT). Computing the inverse of the NDFT is a
  known problem that has been studied extensively in different
  contexts~\cite{ndft1, ndft2}.  Specifically, the NDFT is an
  under-determined system, where the responses of multiple frequency
  elements are unavailable.  Therefore, the inverse of such a Fourier
  transform no longer has a single closed-form solution, but several
  possible solutions.  So how can \name\ pick the best among those
  solutions to find the true times-of-flight of the signal?}

\name\ solves for the inverse-NDFT by adding a constraint to the
inverse-NDFT optimization. Specifically, this constraint favors
solutions that are sparse, i.e., have few dominant paths. Intuitively,
this stems from the fact that while signals in indoor environments
traverse several paths, a few paths tend to dominate as they suffer
minimal attenuation~\cite{channelsparse}.\footnote{We empirically
  evaluate the sparsity of indoor multipath profiles in typical
  line-of-sight and non-line-of-sight settings in~\xref{sec:restof}.}
Indeed other localization systems make this assumption as well, albeit
less explicitly. For instance, antenna-array systems can resolve a
limited number of dominant paths based on the number of antennas they
use.

We can formulate the sparsity constraint mathematically as follows.
Let the vector ${\bf p}$ sample inverse-NDFT at $m$ discrete values
$\tau \in \{\tau_1, \dots, \tau_m\}$. Then, we can introduce sparsity
as a simple constraint in the NDFT inversion problem that minimizes
the L-1 norm of ${\bf p}$. Indeed, it has been well-studied in
optimization theory that minimizing the L-1 norm of a vector favors
sparse solutions for that vector~\cite{l1sparse}.  Thus, we can write
the optimization problem to solve for the inverse-NDFT as:
\begin{align}
\min & \|{\bf p}\|_1\\
\text{s.t.~~~}\|{\bf \tilde{h}} - & {\bf \mathcal{F}p}\|_2^2  = 0
\end{align}
where, ${\bf \mathcal{F}}$ is the $n \times m$ Fourier matrix,
i.e. ${\bf \mathcal{F}}_{i,k}=e^{-j 2\pi f_{i, 0} \tau_k}$, ${\bf
  \tilde{h}} = [\tilde{h}_{1, 0}, \dots, \tilde{h}_{n, 0}]^T$ is the
$n \times 1$ vector of wireless channels at the $n$ different
center-frequencies $\{f_{1,0}, \dots, f_{n,0}\}$, $\|\cdot\|_1$ is the
L-1 norm, and $\|\cdot\|_2$ is the L-2 norm. Here, the constraint
makes sure that the Discrete Fourier Transform of ${\bf p}$ is ${\bf
  \tilde{h}}$, as desired. In other words, it ensures ${\bf p}$ is a
candidate inverse-NDFT solution of ${\bf \tilde{h}}$. The objective
function favors sparse solutions by minimizing the L-1 norm of ${\bf
  p}$.

We can re-formulate the above optimization problem using the method of
Lagrange multipliers as:
\begin{align}
\min_{{\bf p}} \|{\bf \tilde{h}} -{\bf \mathcal{F}p}\|_2^2+\alpha\|{\bf p}\|_1 \label{eqn:min}
\end{align}
Notice that the factor $\alpha$ is a sparsity parameter that enforces
the level of sparsity. A bigger choice of $\alpha$ leads to fewer
non-zero values in ${\bf p}$.

The above objective function is convex but not differentiable.  Our
approach to optimize for it borrows from proximal gradient methods, a
special class of optimization algorithms that have provable
convergence guarantees~\cite{proxconv}.  Specifically, our algorithm
takes as inputs the measured wireless channels ${\bf \tilde{h}}$ at
the frequencies $\{f_{1,0}, \dots, f_{n,0}\}$ and the sparsity
parameter $\alpha$. It then applies a gradient-descent style algorithm
by computing the gradient of differentiable terms in the objective
function (i.e. the L-2 norm), picking sparse solutions along the way
(i.e. enforcing the L-1 norm). Algorithm~\ref{alg:ndft} summarizes
these steps.  \name\ runs this algorithm to invert the NDFT and find
the multipath profile.

\begin{algorithm}
{
\begin{algorithmic}[0]
{ \footnotesize
\State \LComment{Given: Measured Channels, ${\bf \tilde{h}}$ }
\State \LComment{${\bf \mathcal{F}}$: Non-uniform DFT matrix, such that ${\bf \mathcal{F}}_{i,k}=e^{-j2\pi f_{i,0} \tau_k}$}
\State \LComment{$\alpha$: Sparsity parameter; $\epsilon$: Convergence Parameter}
\State \LComment{Output: Inverse-NDFT, ${\bf p}$}
\State \LComment Initialize ${\bf p}_0$ to a random value, $t=0$, $\gamma=\frac{1}{||{\bf \mathcal{F}}||_2}$.
\While{$converged=false$}
	\State {${\bf p}_{t+1}=$\textproc{sparsify}(${\bf p}_t-\gamma {\bf \mathcal{F}}^*({\bf E}{\bf p}_t-{\bf \tilde{h}}),\gamma\alpha$)}
	\If{$||{\bf p}_{t+1}-{\bf p}_{t}||_2<\epsilon$}
		\State	$converged=true$
		\State ${\bf p}={\bf p}_{t+1}$
	\Else
		\State	$t=t+1$
	\EndIf
\EndWhile

\Function{sparsify}{${\bf p}$,$t$}
\For{$i=1,2,...length({\bf p})$}
		\If{$|{\bf p}_i|<t$}
			\State ${\bf p}_i=0$
		\Else
			\State ${\bf p}_i={\bf p}_i\frac{|{\bf p}_i|-t}{|{\bf p}_i|}$
		\EndIf
\EndFor
\EndFunction
}
\end{algorithmic}}
\caption{Algorithm to Compute Inverse NDFT}
\label{alg:ndft}
\end{algorithm}


Finally, \name\ needs to resolve the time-of-flight of the wireless
device based on its multipath profile. To do this, \name\ leverages a
simple observation: Of all the different paths of the wireless signal,
the direct path is the shortest. Hence, the time-of-flight of the
direct path is the propagation delay corresponding to the first peak
in the multipath profile.

We make the following observations: (1) By making the sparsity
assumption, we lose the propagation delays of extremely weak paths in
the multipath profile. However, \name\ only needs the propagation
delay of the direct path. As long as this path is among the dominant
signal paths, \name\ can retrieve it accurately. Of course, in some
unlikely scenarios, the direct path may be too attenuated in the
multipath profile. Like most localization systems, including
angle-of-arrival based approaches, this results in outliers with
poorer localization accuracy. (2) Leveraging sparse recovery of
time-of-flight is key to \name's high resolution. Specifically, sparse
recovery algorithms are well-known to recover sparse useful
information at high resolution, as opposed to all information at low
resolution~\cite{sparsebook}.  Our results in~\xref{sec:restof} depict
the sparsity of representative multipath profiles in line-of-sight and
non-line-of-sight, and show its impact on overall accuracy
in~time-of-flight.

\section{Correcting for Frequency Offsets} \label{sec:practical}
As mentioned in~\xref{sec:tof}, Wi-Fi radios in practice experience
Carrier Frequency Offsets (CFO) that need to be corrected, to apply
\name's algorithms. These offsets occur due to small differences in
the carrier frequency of the transmitting and receiving radio. Such
differences accumulate quickly over time and result in large phase
errors in wireless channel measurements, that must be corrected to
retrieve time-of-flight. We refer to these measured channels from
Wi-Fi radios as channel state information (CSI).

To remove frequency offsets from CSI at the receiver, \name\ exploits
the ACKs that receivers send for every packet from the transmitter
during \name's channel hopping protocol. This means that \name\ can
access another CSI -- this time measured at the transmitter for the
receiver's ACK. This additional CSI is valuable to help mitigate the
frequency offset. To see why, let $f_{i,0}^{tx}$ and $f_{i,0}^{rx}$
denote the center-frequencies of the $i^{\text{th}}$ frequency band of
Wi-Fi at the frequency offset. The frequency offset measured at the
receiver for the transmitter's packet is therefore $f_{i,0}^{rx} -
f_{i,0}^{tx}$. As a result, any phase error in the CSI is proportional
to this offset. In contrast, the frequency offset measured at the
transmitter for the receiver's ACK is $f_{i, 0}^{tx} - f_{i, 0}^{rx}$,
since the transmitter and receiver flip roles. In other words, its
frequency offset is negative of that of the receiver.  As a result,
its measured phase error is also the negative of the phase error at
the receiver. This means that by adding the phases at the receiver and
transmitter (or equivalently, multiplying the CSIs), we can eliminate
any phase error due to frequency offset.

Mathematically, we can observe this property by writing the channel
state information $\text{CSI}_{i, 0}^{rx}(t)$ and $\text{CSI}_{i,
  0}^{tx}(t)$ corrupted by frequency offsets, measured at the receiver
and transmitter respectively, at center-frequencies $f_{i, 0}^{rx}$,
$f_{i, 0}^{tx}$ of Wi-Fi frequency band $i$ at time $t$, as follows:
\begin{align}
{\text{CSI}}_{i, 0}^{rx}(t) = \tilde{h}_{i,0} e^{j (f_{i,0}^{tx} - f_{i,0}^{rx}) t} \label{eqn:hrx}\\
{\text{CSI}}_{i, 0}^{tx}(t) = \tilde{h}_{i,0} \kappa e^{j (f_{i,0}^{rx} - f_{i,0}^{tx}) t} \label{eqn:htx}
\end{align}
Notice that without frequency offsets, the transmitter's channel
equals the receiver's, barring a constant factor $\kappa$ that can be
pre-calibrated. Here, $\kappa$ depends only on the transmit and
receive chains of the device, and is independent of device
location. This is a well-known property of wireless channels called
reciprocity~\cite{reciprocity}. We can therefore multiply the above
equations to recover~the~wireless~channel as follows:
\begin{align}
\tilde{h}_{i,0}^2 = \frac{1}{\kappa} {\text{CSI}}_{i, 0}^{rx}(t) {\text{CSI}}_{i, 0}^{tx}(t) \label{eqn:hrecip}
\end{align}

Of course, the above formulation helps us only retrieve the square of
the wireless channels $\tilde{h}_{i,0}^2$.  However, this is not an
issue: \name\ can directly feed $\tilde{h}_{i,0}^2$ into its algorithm
(Alg.~1 in~\xref{sec:multipath}) instead of $\tilde{h}_{i,0}$. Then
the first peak of the resulting multipath profile will simply be at
twice the time-of-flight.

To see why, let us look at a simple example. Consider a transmitter
and receiver obtaining their signals along two paths, with propagation
delays 2~ns and 4~ns. We can write the square of the resulting
wireless channels from Eqn.~\ref{eq:h} for frequency band $i$ in a
simple form:
\begin{align*}
\tilde{h}_{i, 0}^2 & = (a_1 e^{-j 2\pi f_{i, 0}\times  2} + a_2 e^{-j 2\pi f_{i, 0}\times  4})^2 \\
                   & = a_1^2 e^{-j 2\pi f_{i, 0}\times  2\times 2} + 2 a_1 a_2 e^{-j 2\pi f_{i, 0}\times  (2 + 4)}  + a_2^2 e^{-j 2\pi f_{i, 0}\times  4\times 2} \\
                   & = b_1 e^{-j 2\pi f_{i, 0}\times  4} + b_2 e^{-j 2\pi f_{i, 0}\times  6}  + b_3 e^{-j 2\pi f_{i, 0}\times  8} 
\end{align*}
Where $b_1=a_1^2$, $b_2=2a_1a_2$, $b_3=a_2^2$. Clearly, the above
equation has a form similar to a wireless channel with propagation
delays 4~ns, 6~ns and 8~ns respectively. This means that applying
\name's algorithm will result in peaks precisely at 4~ns, 6~ns and
8~ns. Notice that in addition to 4~ns and 8~ns that are simply twice
the propagation delays of genuine paths, there is an extra peak at
6~ns. This peak stems from the square operation in $\tilde{h}_{i,
  0}^2$ and is a sum of two delays. However, the sum of any two delays
will always be higher than twice the lowest delay. Consequently, the
smallest of these propagation delays is still at 4~ns --- i.e., at
twice the time-of-flight. A similar argument holds for larger number
of signal paths, and can be used to recover time-of-flight.

We make a few important observations: (1) In practice, the forward and
reverse channels cannot be measured at exactly the same $t$ but within
short time separations (tens of microseconds), resulting in a small
phase error. However, this error is significantly smaller than the
error from not compensating for frequency offsets altogether (for tens
of milliseconds). The error can be resolved by averaging over several
packets. (2) Constants such as $\kappa$ and other delays in
transmit/receive hardware result in a constant error in
time-of-flight. These can be pre-calibrated a priori and only once by
measuring time-of-flight to a device at a known distance.

\begin{figure}
\centering		
\includegraphics[width=0.37\textwidth, height=0.17\textheight]{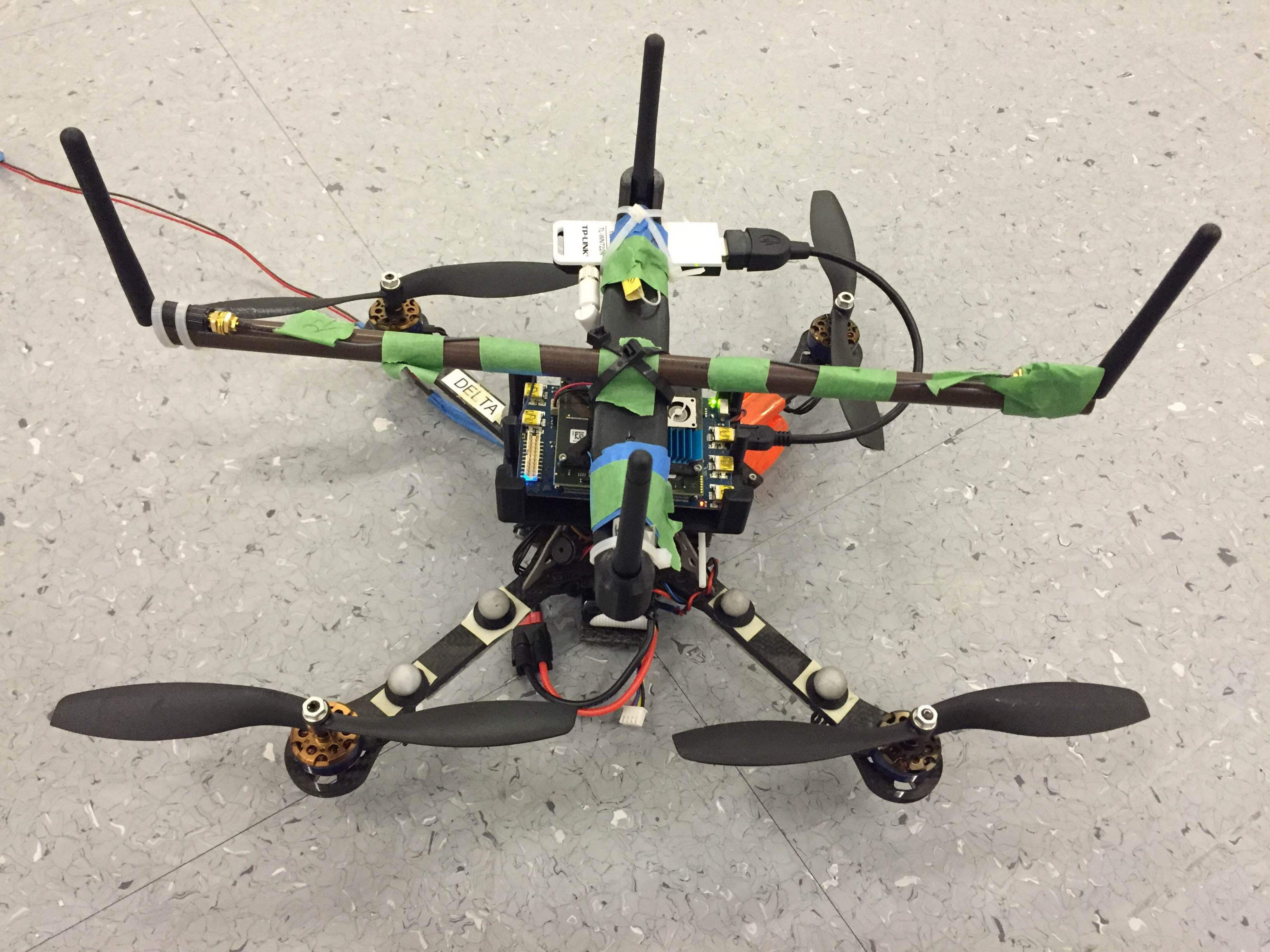}
\caption{\footnotesize{\bf \name\ Personal Drone:} We implement
  \name\ on an AscTec Hummingbird quadrotor with an AscTec
  Atomboard. }
\vspace*{-0.0in}
\label{fig:droneimpl}
\vspace*{-0.1in}
\end{figure}

\section{Computing Distances and Location}\label{sec:localization}
So far, we have explained how \name\ measures the time-of-flight
between two antennas on a pair of Wi-Fi cards. One can then compute
the distance between the two antennas (i.e., the two devices) by
multiplying the time-of-flight by the speed of light. One can also
compute the location by intersecting multiple such distances.

For example, consider a two-antenna receiver that aims to compute its
location relative to a single-antenna transmitter. The receiver first
applies \name's algorithm to measure the time-of-flight of the
transmitter's signal to its two receive antennas. When multiplied by
the speed of light this provides two distances of the two receive
antennas from the transmitter. Hence, the transmitter must lie at the
intersection of the two circles, centered around each receive antenna
with radii defined by these distances.

In general, two distances are not enough to compute the location as
two circles typically intersect at two points.  \name\ can resolve the
ambiguity using one of two strategies: (1) If the receiver has a third
antenna, \name\ can use it to find a third circle on which the
transmitter should lie. The three circles will together intersect at a
unique point (assuming the antennas are not co-linear). Notice that if
the three circles do not intersect exactly (e.g., due to noise),
\name\ can use well-known least-squares optimizations to pick the
point closest to the three circles~\cite{ubicarse}. Similarly, if the
transmitter has more than one antenna, \name\ can improve the localization accuracy by
computing pairwise distances between the transmit and receive antennas and then, incorporating them 
in the optimization problem.
(2) A second approach to remove location ambiguity leverages
mobility.  A receiver can move towards what it believes to be the
transmitter's location, and re-run \name's algorithm. If the
transmitter indeed moved closer, the chosen location is correct. If
not, one must pick the other possible location of the transmitter. We
employ this strategy to disambiguate the transmitter's location for
the personal drone in~\xref{sec:drone}.

\section{Application to Personal Drones} \label{sec:drone}
To illustrate \name's capabilities, we apply it to indoor personal
drones~\cite{personal}. These drones can follow users around while
maintaining a convenient distance relative to the mobile device in the
user's pocket.  Knowing the distance to the user allows the drone to
take clear optimal pictures by ensuring that the user is within the
frame of view at the right level of zoom.  Users can leverage these drones
to take pictures or videos of them while they are performing an
activity, even in indoor settings where GPS is unavailable.

This application highlights \name's unique benefits:
\begin{Itemize}  
\item {\it Device-to-device solution:} A key feature of \name\ is its
  ability to deliver device-to-device localization -- i.e., enabling
  devices with commercial Wi-Fi cards to accurately localize each
  other without support from surrounding infrastructure. Thus,
  \name\ requires only a Wi-Fi enabled drone and a Wi-Fi device on the
  user. The user may use his personal drone to record his activities
  anywhere, whether at home, at work or in the gym, without requiring
  the access points in these buildings to support localization.
\item {\it Uses commercial Wi-Fi cards:} Indoor drones can carry only
  limited payload for stable flight over long durations. In other
  words, drones simply cannot carry state-of-the-art accurate
  localization hardware such as antenna arrays. Fortunately, since
  \name\ is compatible with commodity Wi-Fi cards, it is possible
  integrate it with a light-weight computing module that weighs 90
  grams and can be carried by small indoor drones.
\end{Itemize}

We built \name\ over an AscTec Hummingbird quadrotor equipped with a
Go-pro camera, as shown in Fig.~\ref{fig:droneimpl}.  To localize the
quadrotor, \name\ uses a 3-antenna Wi-Fi radio and intersects the
distances of the user's device to its 3-antennas.  The distance
measurements are integrated with drone navigation using a standard
negative feedback-loop robotic controller~\cite{loop}. Specifically,
this controller measures the current distance of the user's mobile
device. If the user is closer than expected, the drone takes a
discrete step further away and vice-versa. Such controllers are
well-known to converge efficiently to stable
solutions~\cite{loop}. Our results in~\xref{sec:droneresults} show
that \name\ converges to optimal locations that maintain stable
distances. Further, our approach also benefits from an inherent
synergy between \name's localization and the robotic
controller. Specifically, the feedback controller invokes \name's
algorithm multiple times to compute its precise distance to the
user. In doing so, it can average across these invocations and reject
outliers to maintain this distance at a much higher accuracy than
\name's native algorithm, as we show in~\xref{sec:droneresults}.

\section{Limitations and Trade-Offs}\label{sec:limitations}

In this section, we discuss the limitations and trade-offs in \name's
design.

\vspace*{0.02in}\noindent \textbf{Frequency Band Hopping: }
\name\ requires wireless devices to hop between Wi-Fi frequency
bands. Our implementation hops between all bands of Wi-Fi in 84~ms
(see~\xref{sec:tcpeffect}). A natural question to ask is how this
hopping affects data traffic and user experience. Note that \name\ is
primarily targeted for localization between a pair of Wi-Fi user
devices that may otherwise not exchange data. However, some users may
be interested in running \name\ on a single access point in home
environments, where there may not be multiple access points covering
the same physical space. Such access points cannot transmit/receive
data to other clients as they localize. \xref{sec:tcpeffect} shows
that occasional demands for localization every tens of seconds
minimally impacts TCP and video applications on these clients. But
more frequent requests for localization may necessitate deploying a
dedicated \name\ access point exclusively for in-home
localization. Finally, since \name\ sends few packets per frequency
band, it does not significantly impact nearby~Wi-Fi~networks.

\vspace*{0.02in}\noindent \textbf{Antenna Separation: } \name's
accuracy in localizing a device improves with greater separation
between its receive antennas. As the separation between a pair of
antennas becomes larger, the resulting localization circles experience
smaller overlap. This means that their point of intersection is less
sensitive to noise, improving localization accuracy. Consequently,
\name's positioning accuracy on a Wi-Fi access point which can afford
larger separation between antennas is higher than \name\ between a
pair of user devices (e.g. laptops or tablets). Our results
in~\xref{sec:resloc} empirically evaluate this trade-off in typical
indoor environments.

\begin{figure}
\centering		
\includegraphics[width=0.42\textwidth, height=0.20\textheight]{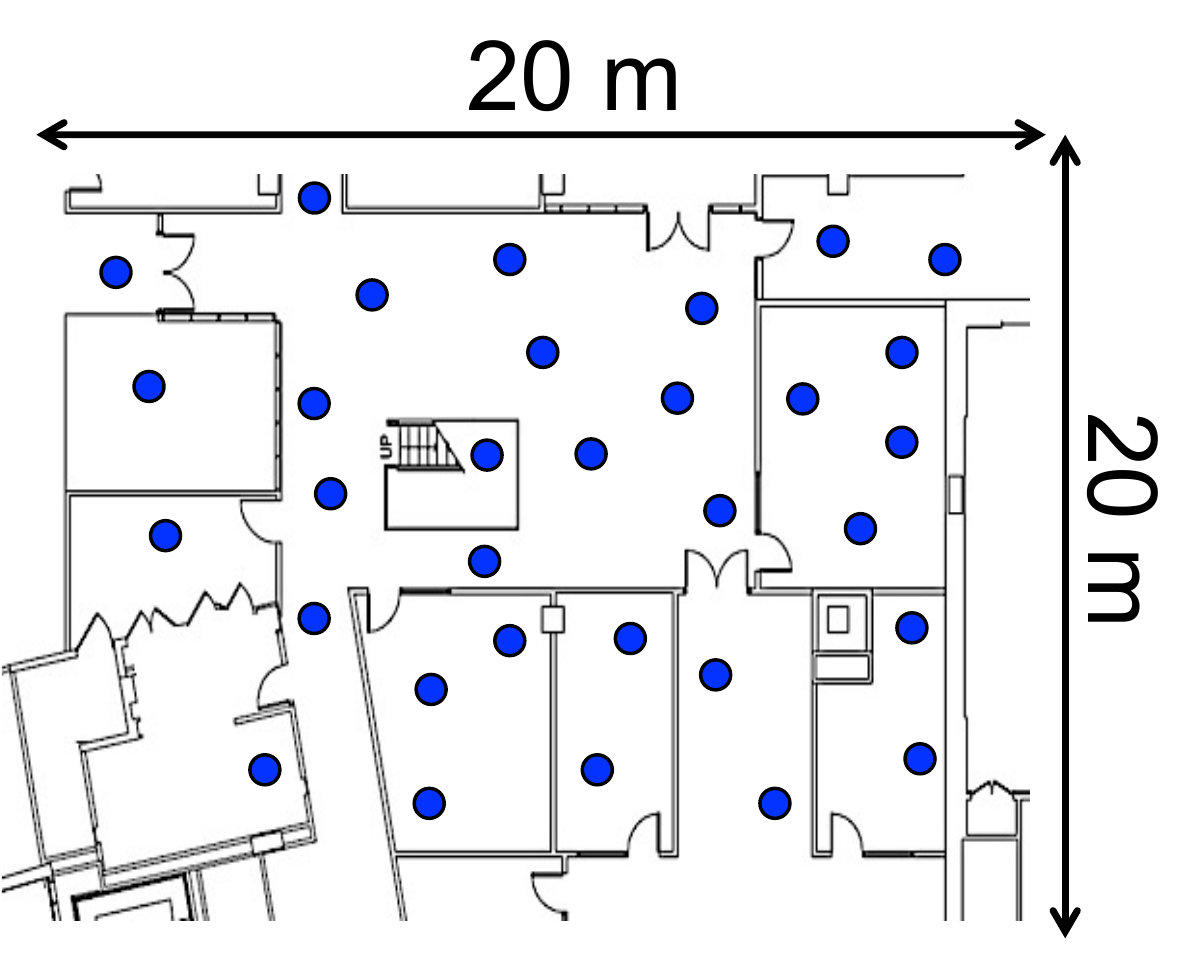}
\vspace*{-0.1in}
\caption{\footnotesize{\bf Testbed:} Blue dots show candidate device
  locations. }
\label{fig:testbed}
\vspace*{-0.15in}
\end{figure}

\begin{figure*}
\centering		
\subfigure[Time of Flight]{\label{fig:toflos}\includegraphics[width=0.32\textwidth,height=0.16\textheight]{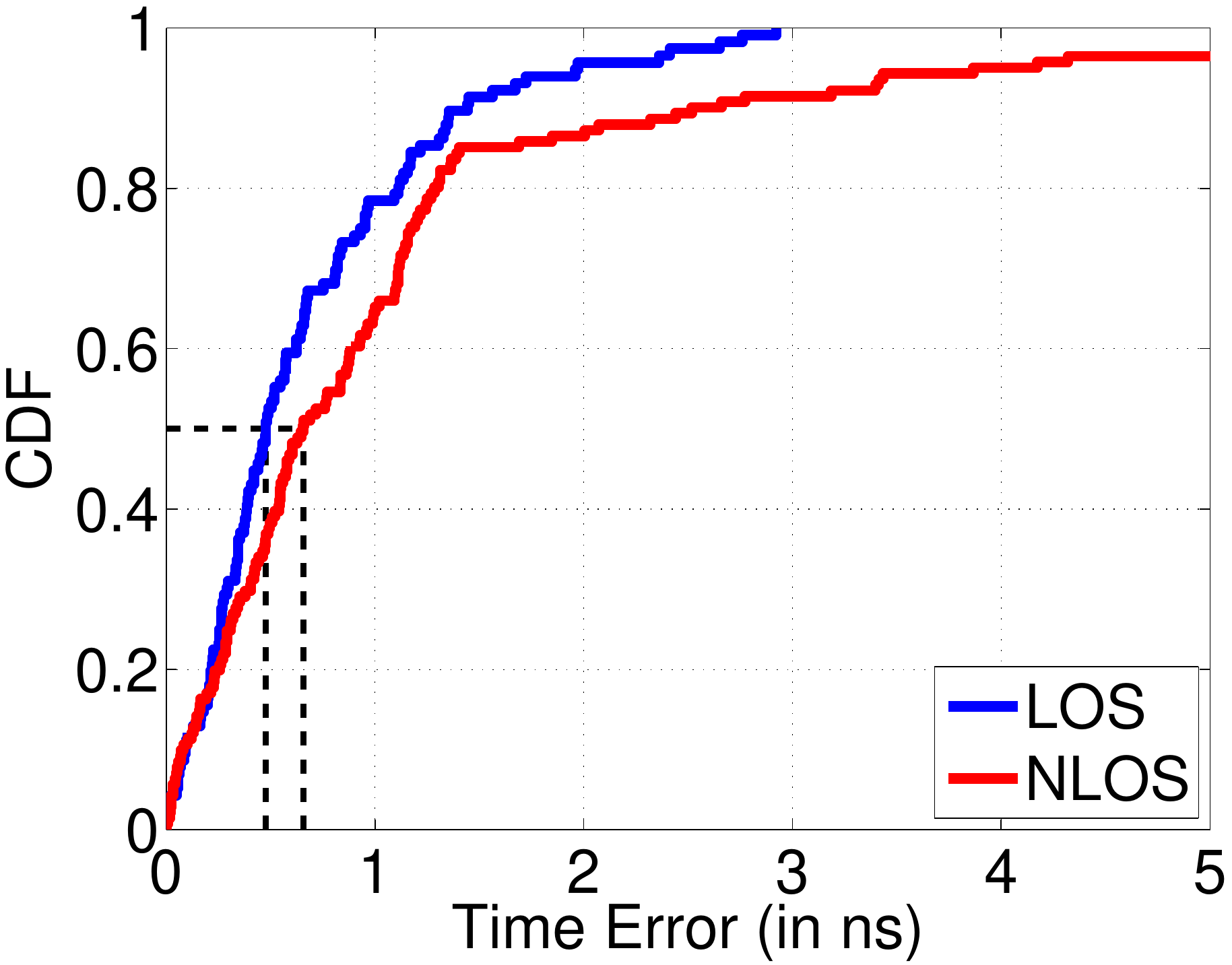}} 
\subfigure[Multipath Profiles]{\label{fig:tofprofile}
\includegraphics[width=0.32\textwidth,height=0.16\textheight]{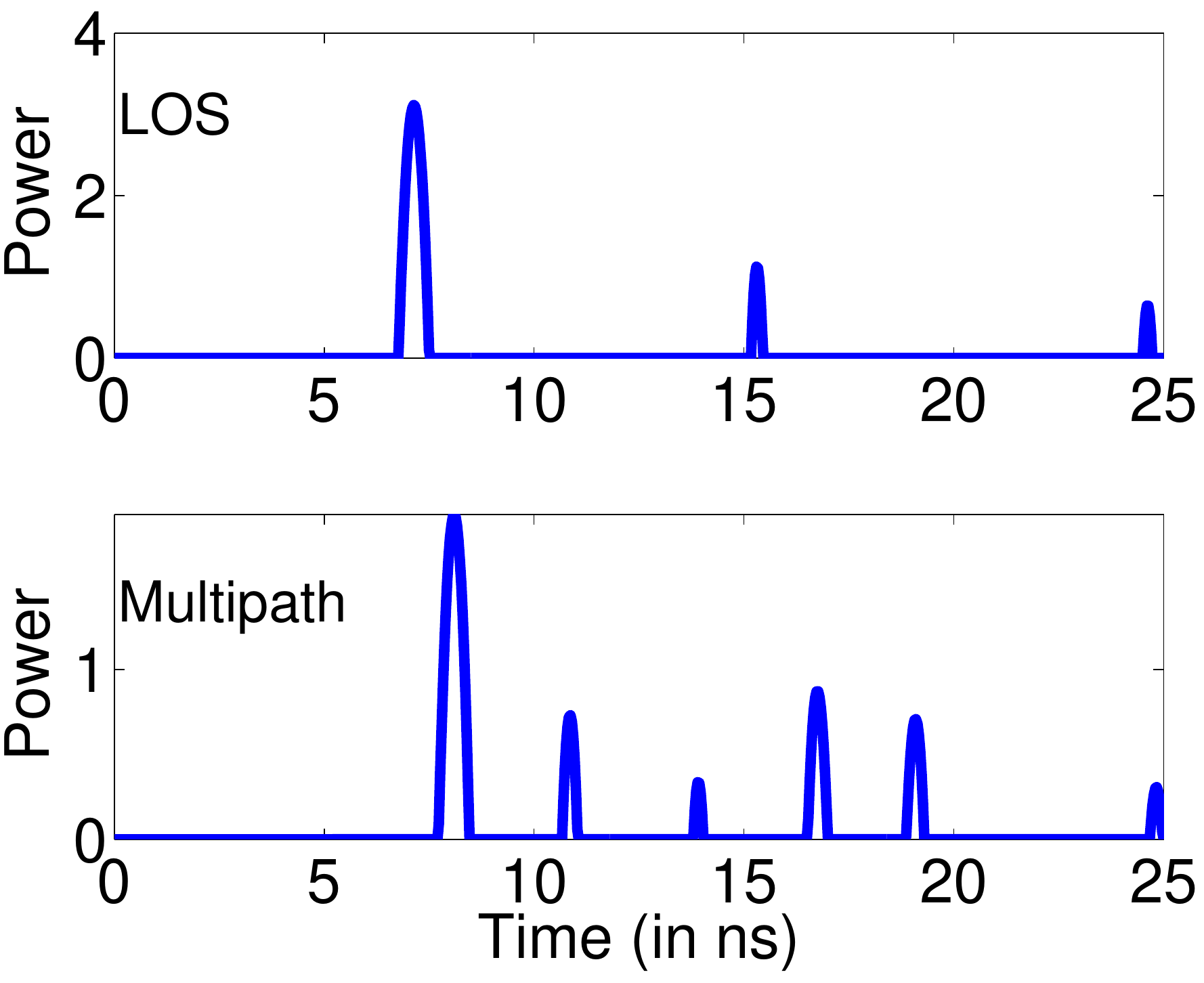}
}
\subfigure[Packet Detection Delay]{\label{fig:resdelay}\includegraphics[width=0.32\textwidth,height=0.16\textheight]{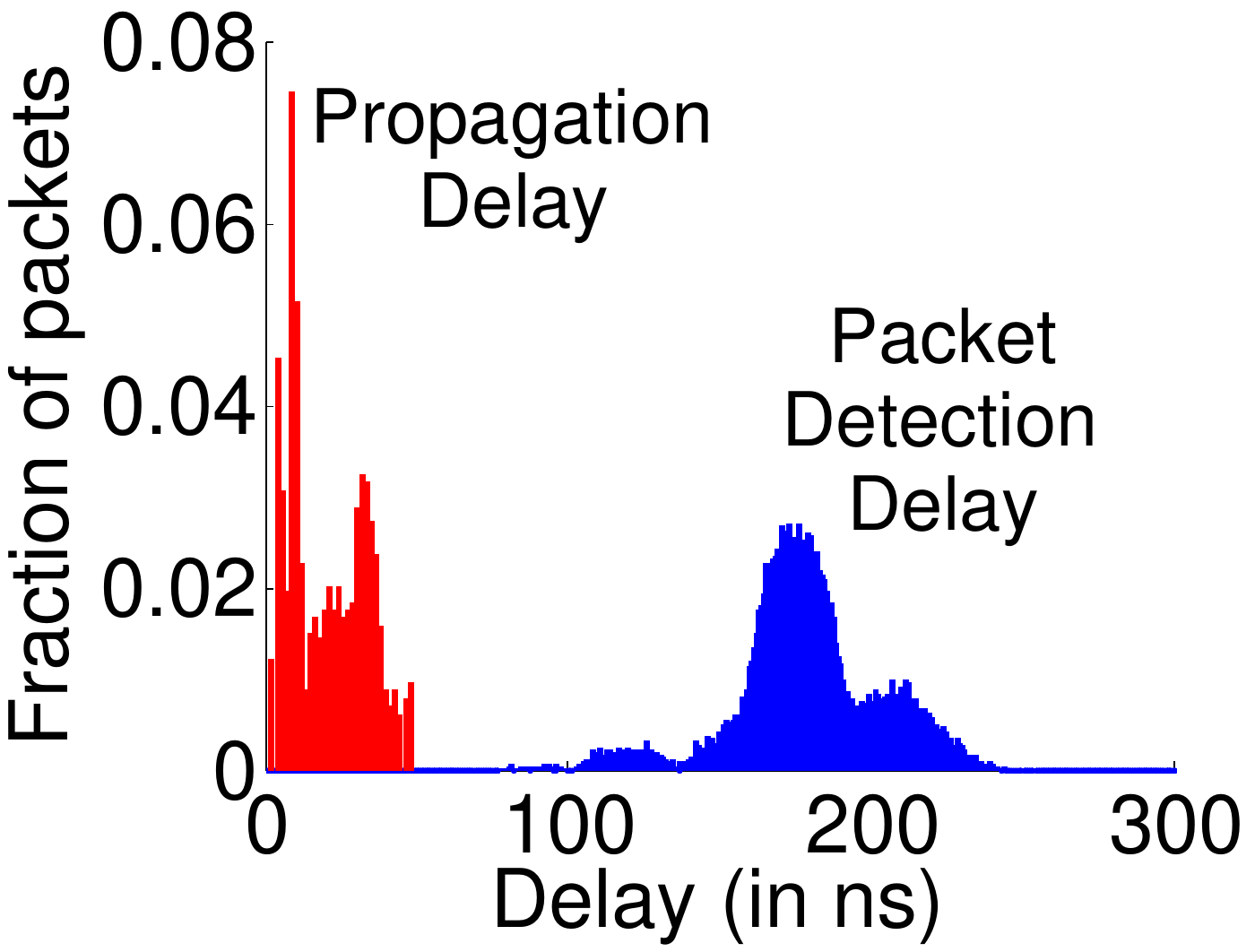}} 
\vspace*{-0.15in}\caption{\footnotesize {\bf Accuracy in Time of
    Flight}: (a) measures the CDF of error in time-of-flight between
  two devices in Line of Sight (LOS) and Non-Line of Sight (NLOS). (b)
  plots representative multipath profiles. (c) plots histograms of
  time-of-flight and packet detection delay. }
\label{fig:tof}
\vspace*{-0.1in}
\end{figure*}

\section{Implementation} \label{sec:implementation}
We implemented \name\ as a software patch to the iwlwifi driver on
Ubuntu Linux running the 3.5.7 kernel. To measure
channel-state-information, we leverage the 802.11 CSI
Tool~\cite{csitool} for the Intel 5300 Wi-Fi card.  We measure
wireless channels on both 2.4~GHz and 5~GHz Wi-Fi bands.\footnote{The
  Intel 5300 Wi-Fi card is known to have a firmware issue on the
  2.4~GHz bands that causes it to report the phase of the channel
  $\angle \tilde{h}_{i,0}$ modulo $\pi/2$ (instead of the phase modulo
  $2\pi$)~\cite{phaser}. We resolve this issue by performing \name's
  algorithm at 2.4~GHz on $\tilde{h}_{i,0}^4$ instead of
  $\tilde{h}_{i,0}$.  This does not affect the fact that the direct
  path of the signal will continue being the first peak in the inverse
  NDFT (like in~\xref{sec:practical}).}

Unless specified otherwise, we pair two \name\ devices by placing each
device in monitor mode with packet injection support on the same Wi-Fi
frequency. We implemented \name's channel hopping protocol
(see~\xref{sec:tof}) in the iwlwifi driver using high resolution
timers (hrtimers), which can schedule kernel tasks such as packet
transmits at microsecond granularity. Since the 802.11 CSI Tool does
not report channel state information for Link-Layer ACKs received by
the card, we use packet-injection to create and transmit special
acknowledgments directly from the iwlwifi driver to minimize delay
between packets and acknowledgments. These acknowledgments are also
used to signal the next channel that the devices should hop to, as
described in~\xref{sec:tof}. Finally we process the channel state
information to infer time-of-flight and device locations purely in
software written in part in C++, MEX and MATLAB.

\section{Results}\label{sec:results}
We evaluate \name\ using the testbed in Fig.~\ref{fig:testbed}.

\subsection{Accuracy in Time-of-Flight} \label{sec:restof} \vspace*{-0.1in}
In this experiment, we evaluate whether \name\ can deliver on its
promise of measuring sub-nanosecond time-of-flight between a single
pair of commodity Wi-Fi devices.

\sssection{Method: } We conduct our main experiments in a floor of a
large office building measuring $20~\text{m}\times 20~\text{m}$ as
shown in Fig.~\ref{fig:testbed}. The floor has multiple offices, a
lounge area, conference rooms, metal cabinets, computers and
furniture. We perform our experiments using two Thinkpad W300 Laptops
equipped with 3-antenna Intel 5300 Wi-Fi cards.  We placed the two
devices randomly at any of 30 randomly chosen locations, as shown by
the blue circles in the figure, with their pairwise distance up to
$15$~m. We perform experiments for pairs of locations both in
line-of-sight and non-line-of-sight. We measure the ground-truth of
these locations using a combination of architectural drawings of our
buildings and a Bosch GLM50 laser distance measurement
tool~\cite{bosch}, which measures distances up to 50~m with an
accuracy of 1.5~mm. We repeat the experiment multiple times and
measure the time-of-flight in each instance. We also compute the
packet-detection delay of each packet using channel phase
(see~\xref{sec:pdelay}) to gauge its effect on the measurement of
time-of-flight.

\sssection{Time-of-Flight Results: } We first evaluate \name's
accuracy in time-of-flight. Fig.~\ref{fig:toflos} depicts the CDF of
the time-of-flight of the signal in line-of-sight settings and
non-line-of-sight.  We observe that the median errors in
time-of-flight estimation are 0.47~ns and 0.69~ns respectively
(95$^{\text{th}}$ percentile: 1.96~ns and 4.01~ns). {Our results show
  that \name\ achieves its promise of computing time-of-flight at
  sub-nanosecond accuracy. To put this in perspective, consider
  SourceSync~\cite{sourcesync}, a state-of-the-art system for time
  synchronization. SourceSync achieves 95$^{\text{th}}$ percentile
  synchronization error up to 20~ns, using advanced software
  radios. In contrast, \name\ achieves order-of-magnitude lower error
  in time-of-flight using commodity Wi-Fi cards. However, we point out
  that unlike indoor positioning, tens of nanoseconds of error is
  sufficient for time-synchronization, which is the application
  SourceSync targets. }

\begin{figure*}
\centering		
\subfigure[Accuracy with distance]{\label{fig:tofdist}\includegraphics[width=0.32\textwidth,height=0.16\textheight]{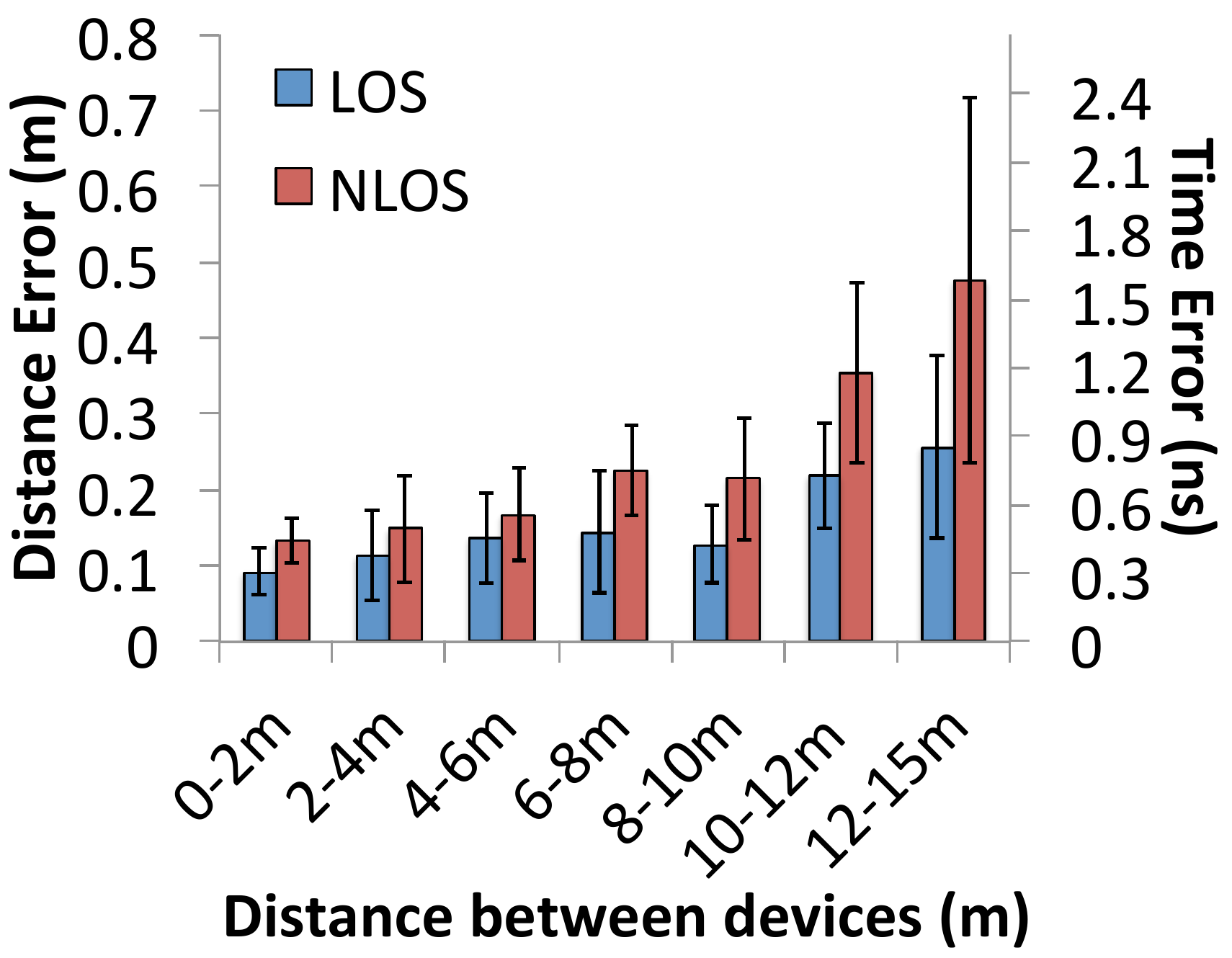}}
\subfigure[Localization Error (Small separation)]{\label{fig:locsmall}\includegraphics[width=0.32\textwidth,height=0.16\textheight]{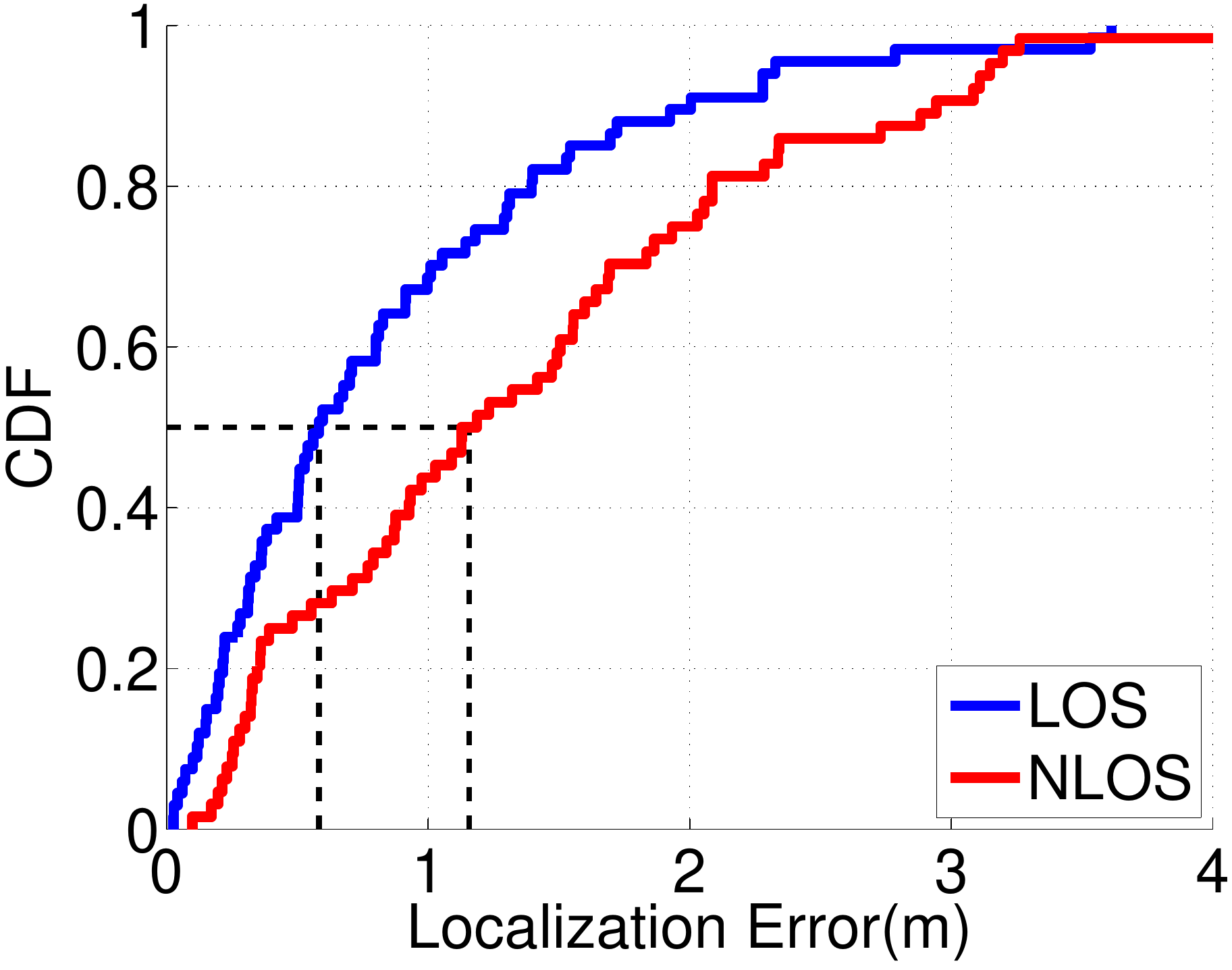}} 
\subfigure[Localization Error (Large separation)]{\label{fig:locbig}\includegraphics[width=0.32\textwidth,height=0.16\textheight]{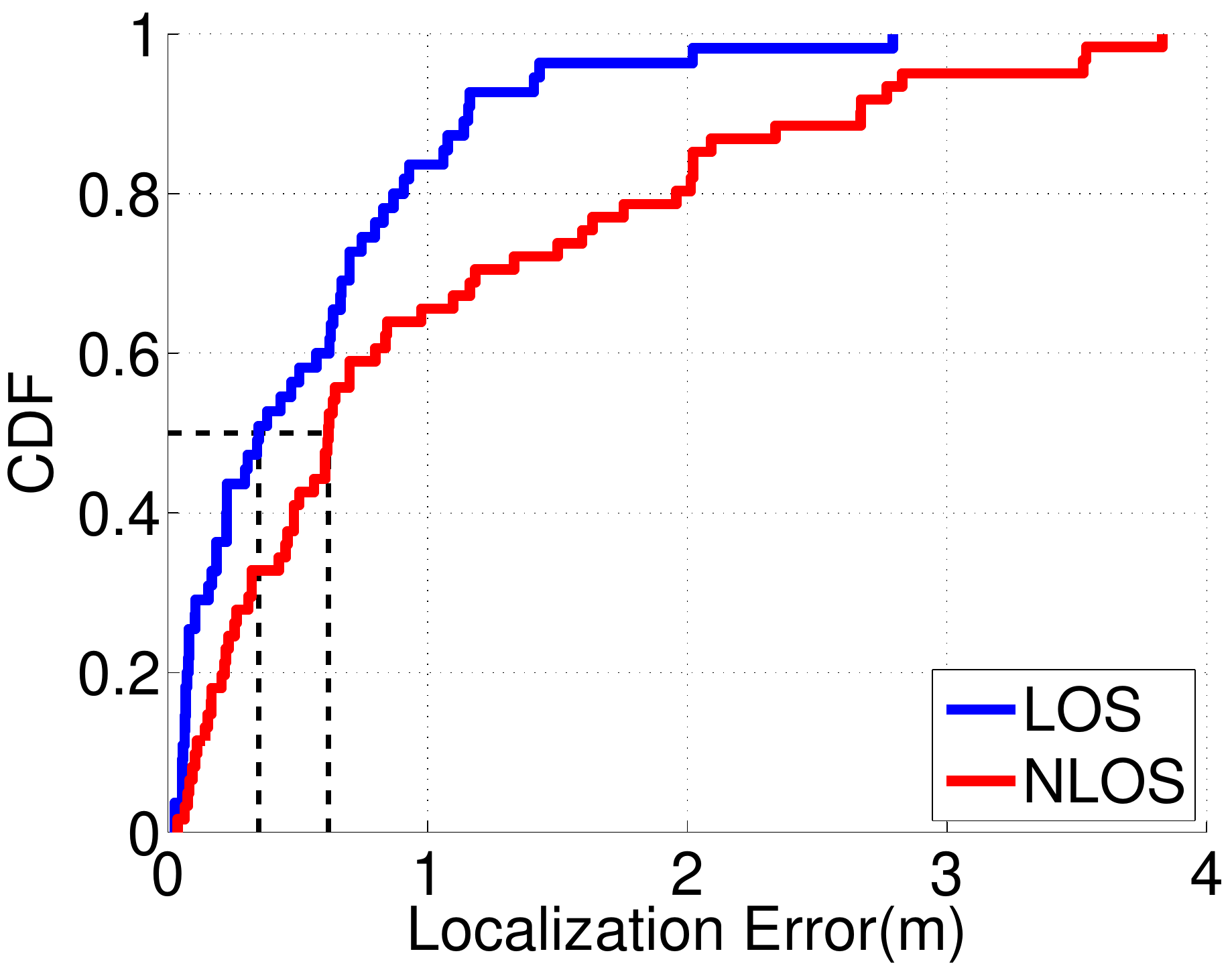}} 
\vspace*{-0.15in}\caption{\footnotesize {\bf (a) Accuracy with Distance:}
  Plots the error in distance/time across the ground-truth distance
  between transmitter and receiver.  {\bf (b, c) Localization Accuracy:}
  Plots CDF of localization error using (b) a client with mean antenna
  separation of 30~cm and (c) an access point with antenna separation
  of 100~cm for transmitter and receiver in Line-of-Sight (LOS) and
  Non-Line-of-Sight (NLOS).}
\label{fig:loc}
\vspace*{-0.0in}
\end{figure*}

\sssection{Multipath Profile Results: } Next, we plot candidate
multipath profiles computed by \name.  Fig.~\ref{fig:tofprofile} plots
representative multipath profiles in line-of-sight and multipath
environments. We note that both profiles are sparse, with the profile
in multipath environments having five dominant peaks. Across
experiments, the mean number of dominant peaks in the multipath
profiles is 5.05 on average, with standard deviation 1.95 ---
indicating that they are indeed sparse. As expected, the profile in
line-of-sight has even fewer dominant peaks than the profile in
multipath settings. In both cases, we observe that the leftmost peaks
in both the profiles correspond to the true location of the source.
Further, we observe that the peaks in both profiles are sharp due to
two reasons: 1) \name\ effectively spans a large bandwidth that
includes all Wi-Fi frequency bands, leading to high time resolution;
2) \name's resolution is further improved by exploiting sparsity that
focuses on retrieving the sparse dominant peaks at much higher
resolution, as opposed to all peaks.

\sssection{Packet Detection Delay Results: } We compare time-of-flight
in indoor environments against packet detection
delay. Fig.~\ref{fig:resdelay} depicts histograms of both packet
detection delay and time-of-flight across experiments. \name\ observes
a median packet detection delay of 177~ns across experiments. We
emphasize two key observations: (1) Packet detection delay is nearly
8$\times$ larger than the time-of-flight in our typical indoor
testbed. (2) It varies dramatically between packets, with a high
standard deviation of 24.76~ns. In other words, packet detection delay
is a large contributor to time-of-arrival that is highly variable, and
therefore, hard to predict. This means that if left uncompensated,
these delays could lead to a large error in time-of-flight
measurements. Our results therefore reinforce the importance of
accounting for these delays and demonstrate \name's ability to do so.

\sssection{Distance Results: } Fig.~\ref{fig:tofdist} plots the median
and standard deviation of error in distance computed between the
transmitter and receiver against their true relative distance. We
observe that this error is initially around 10~cm and increases to at
most 25.6~cm at 12-15 meters. The increase is primarily due to reduced
signal-to-noise ratio at further distances.
\vspace*{-0.02in}

\subsection{Localization Accuracy} \label{sec:resloc}
Next, we evaluate \name's accuracy in finding the indoor position of
one device relative to another.

\sssection{Method: } We repeat the experiment for the setup
in~\xref{sec:restof} using a pair of 3-antenna client laptops with
antennas separated by a mean distance of 30~cm. We consider pairs of
locations where the distance between the devices vary up to 15~m. We
then measure the time-of-flight of the transmitter's signal to each
antenna of the receiver. We multiply this quantity by the speed of
light to measure the pairwise distances between the antennas on the
transmitter and the receiver. We perform outlier rejection on this set
of distance estimates to discard estimates that do not fit the
geometry of the relative antenna placements on these devices. Next, we
use the remaining distance estimates to compute the location of the
device using a least-square optimization formulation (as stated in
\xref{sec:localization}). We repeat the experiment multiple times in
line-of-sight and non-line-of-sight.

\sssection{Results: } Fig. \ref{fig:locsmall} plots a CDF of
localization error using \name\ in different settings. The device's
median positioning error for line of sight scenarios is 58~cm and
118~cm in line-of-sight and non-line-of-sight. Thus, \name\ achieves
state-of-the-art indoor localization accuracy between a pair of user
devices without third party support.

As mentioned in~\xref{sec:limitations}, \name's accuracy depends on
the separation between antennas. In particular, users may wish to run
\name\ to localize their device relative to the single Wi-Fi access
point in their home, where multiple access points covering the same
area may be unavailable. Such an access point can afford greater
separation between antennas than a user device. To evaluate this, we
repeated the above experiment with the receiving laptop emulating a
Wi-Fi access point with antennas separated by 100~cm. In this setting,
the median localization error, reduces as expected to 35~cm and 62~cm
in line-of-sight and non-line-of-sight (see Fig. \ref{fig:locbig}).

\begin{figure*}
\vspace*{-0.05in}
\centering		
\subfigure[Hopping Time]{\label{fig:hoptime}\includegraphics[width=0.32\textwidth,height=0.16\textheight]{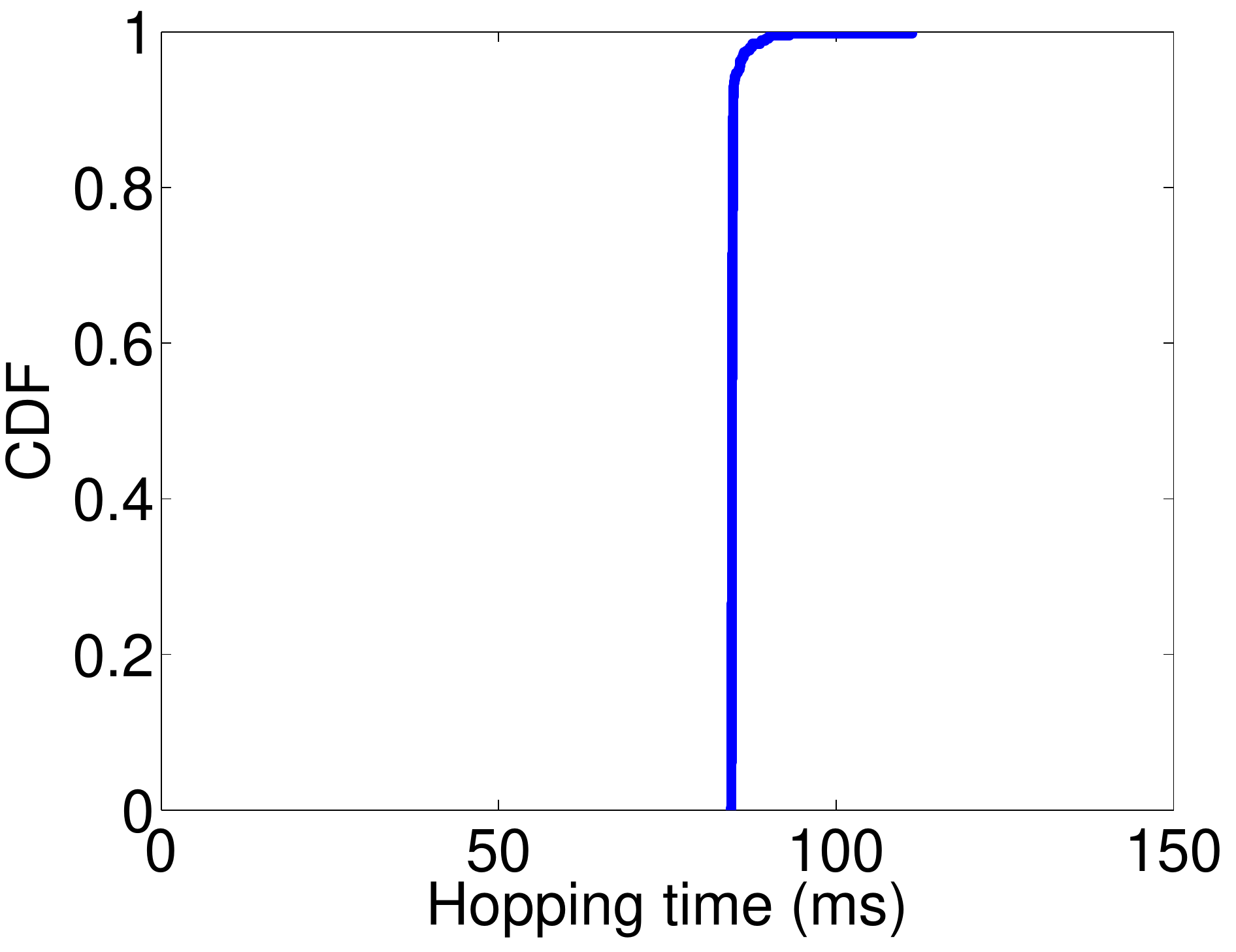}} 
\subfigure[Video Streaming]{\label{fig:viddel}\includegraphics[width=0.32\textwidth,height=0.16\textheight]{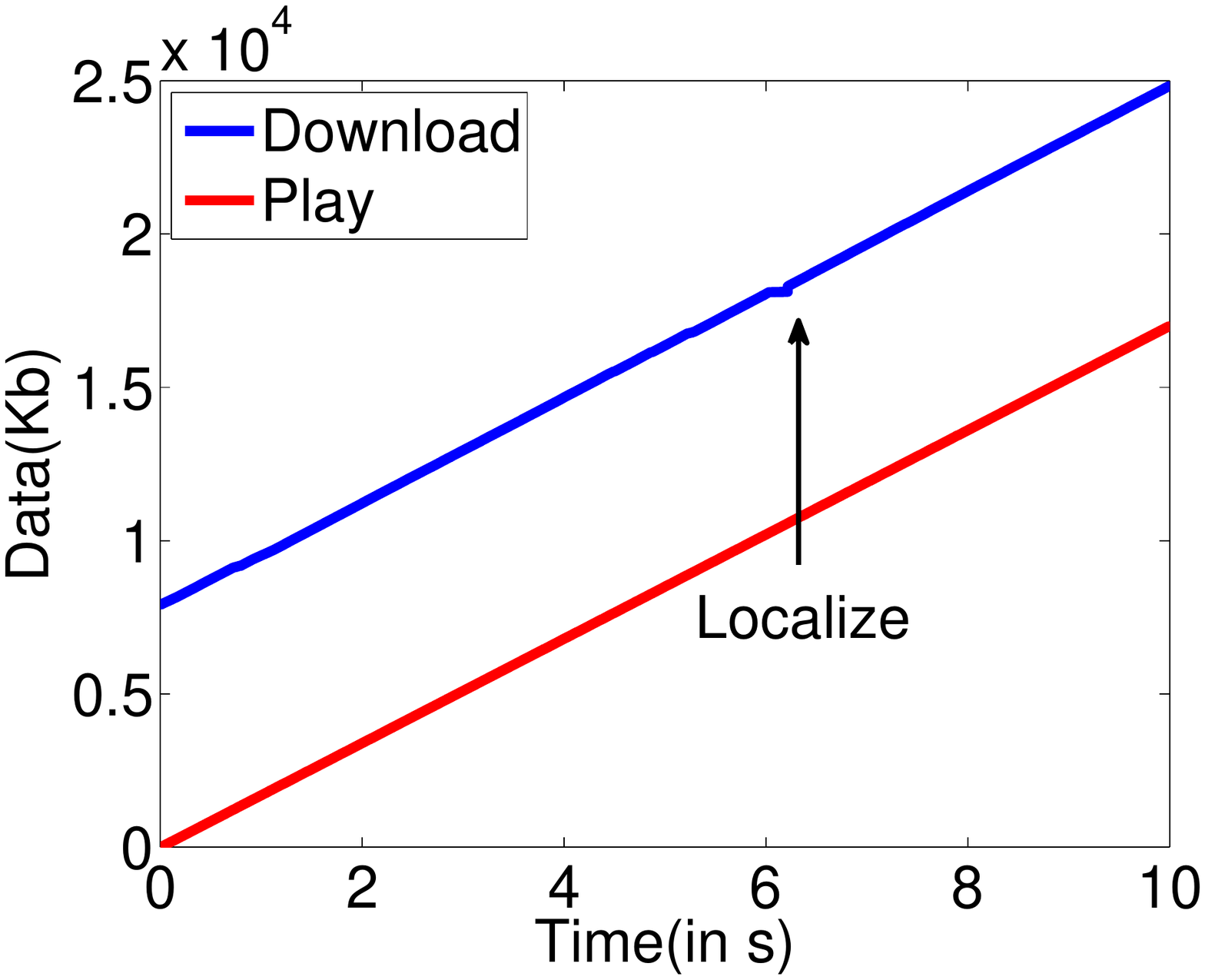}} 
\subfigure[TCP Throughput]{\label{fig:tcptpt}\includegraphics[width=0.32\textwidth,height=0.16\textheight]{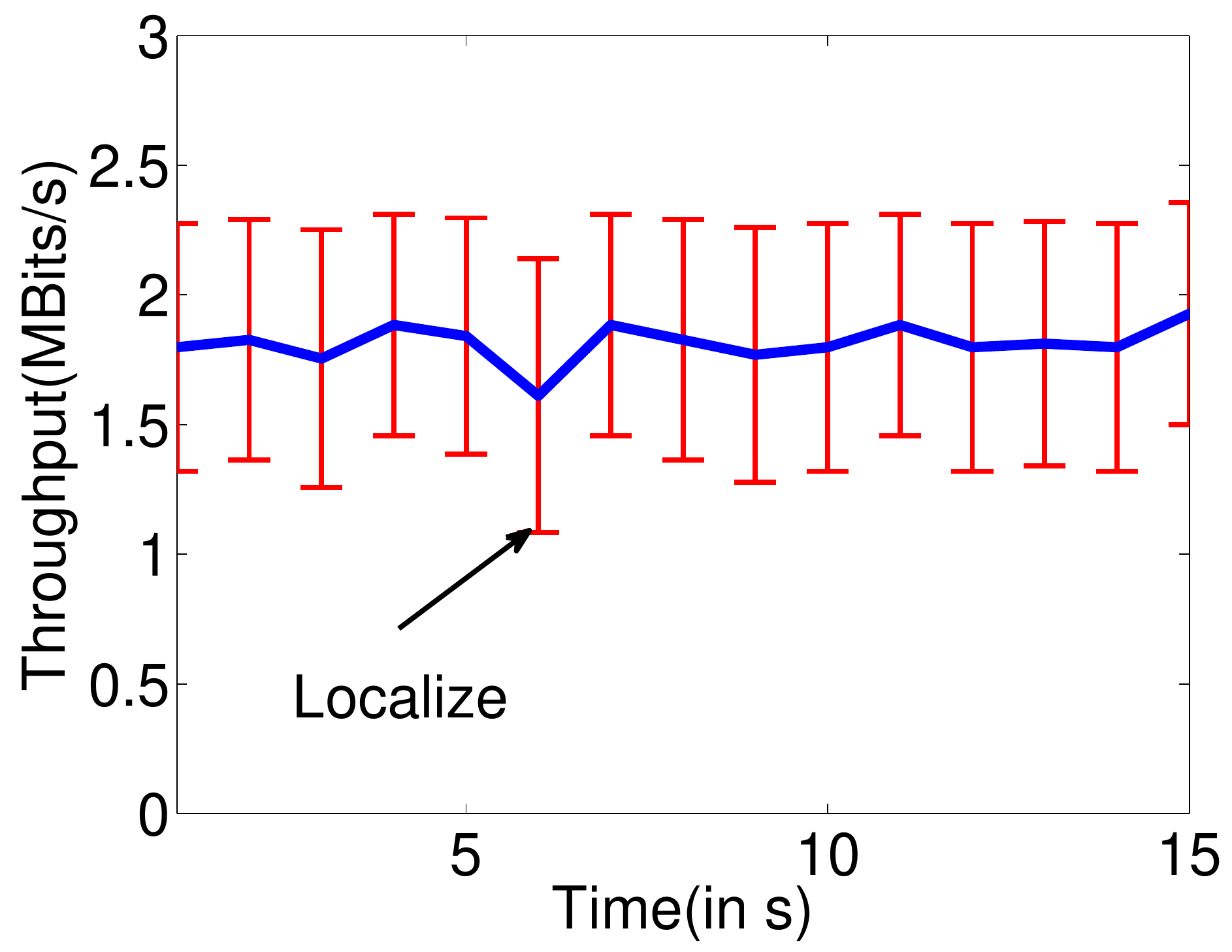}} 
\vspace*{-0.2in}\caption{\footnotesize{{\bf Impact on Network
      Traffic: } (a) measures the CDF time taken by \name\ to hop
    between all Wi-Fi bands -- a small value of 84~ms. Consider a
    client-1 with a long-running traffic flow to an access point. The
    access point is asked to localize another client-2 at $t =
    6$~s. (b) depicts a representative trace of the number of bytes of
    data downloaded and data played over time if the client-1 views a
    VLC video stream. (c) measures the throughput if client-1 runs a
    TCP flow using iperf. In either case, the impact of client-1's
    flow is minimal at $t = 6$~s.}}
\label{fig:traffic}
\vspace*{-0.1in}
\end{figure*}

\begin{figure*}
\centering		
\subfigure[Error in Distance]{\label{fig:dronedist}\includegraphics[width=0.32\textwidth,height=0.16\textheight]{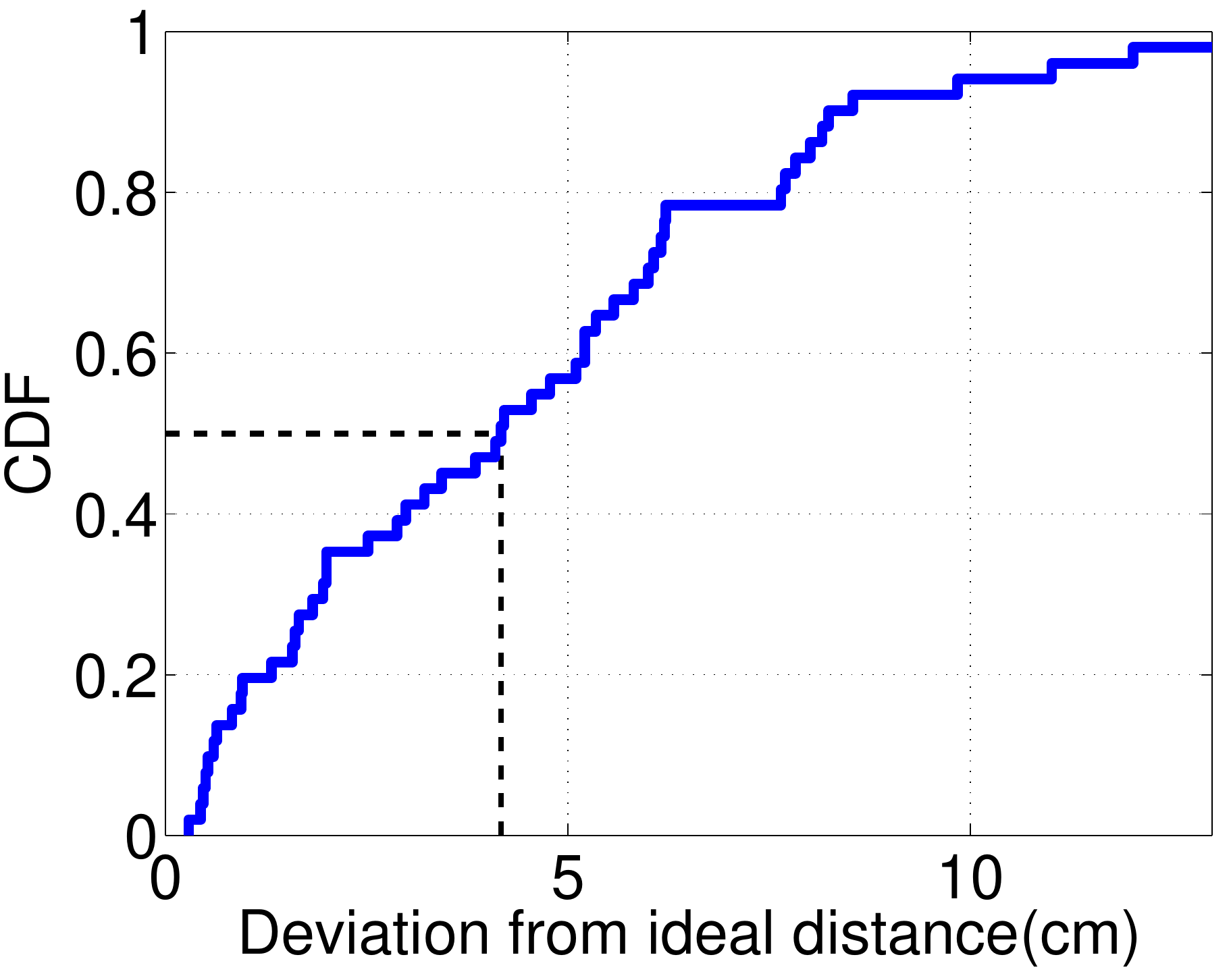}} \hfill \hspace*{-0.2in}
\subfigure[Drone Trajectory]{\label{fig:dronetraj}\includegraphics[width=0.32\textwidth,height=0.172\textheight]{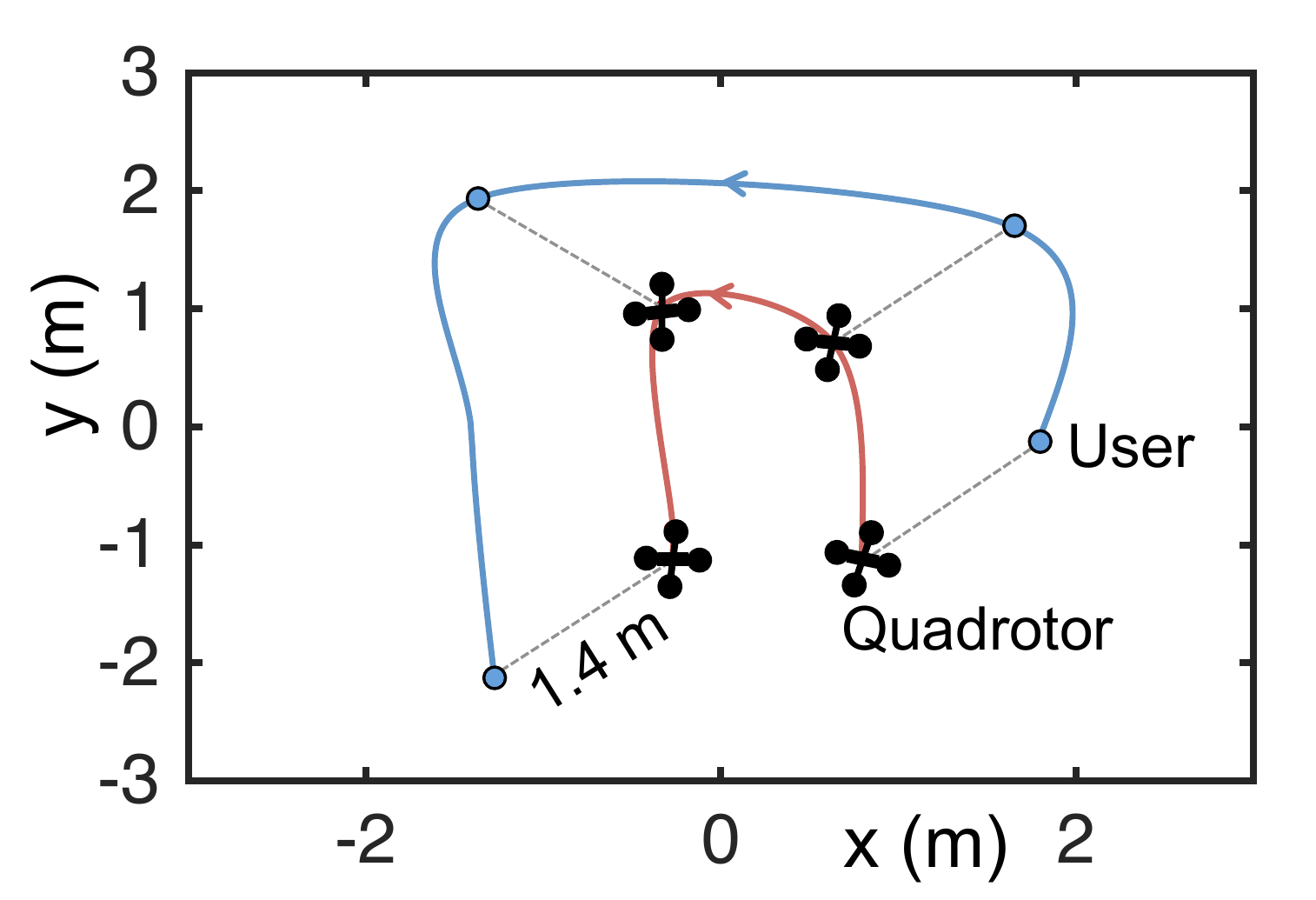}} \hfill
\subfigure[Picture of User]{\label{fig:dronepic}\includegraphics[width=0.28\textwidth,height=0.16\textheight]{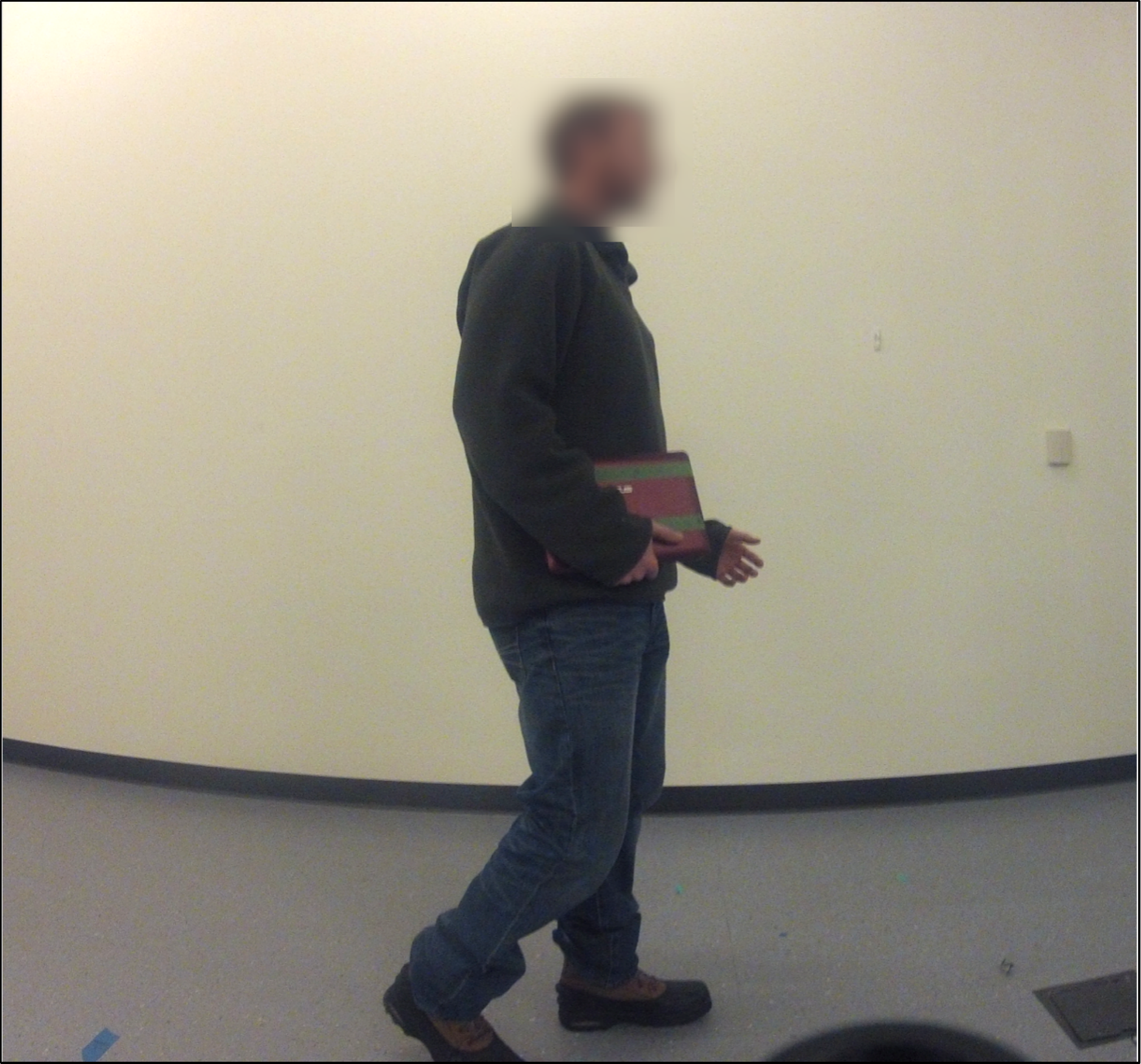}} 
\vspace*{-0.15in}
\caption{\footnotesize{\bf Application to Personal Drones: } The
  personal drone uses \name\ to maintain a constant distance of 1.4~m
  to the user. (a) depicts the CDF of error in distance of the drone
  relative to 1.4~m. (b) depicts a candidate trajectory of the
  drone. (c) shows an example picture of the user (face blurred for
  anonymity) taken by the drone. Since its Go-Pro uses a wide-angle
  lens, the user is fully in-frame at 1.4~m.  }
\label{fig:personaldrone}
\vspace*{-0.15in}
\end{figure*}
\vspace*{-0.1in}
\subsection{Impact on Network traffic}\label{sec:tcpeffect}
\name\ is primarily targeted to enable localization-between a pair of
user devices, which may not otherwise communicate data between each
other directly. However, an interesting question is the impact of
\name\ on network traffic, if one of the devices is indeed serving
traffic, e.g., a Wi-Fi access point.  This experiment answers three
questions in this regard: (1) How long does \name\ take to hop between
all Wi-Fi bands? (2) How does \name\ impact real-time traffic like
video streaming applications? (3) How does \name\ affect TCP? We
address these questions below:

\sssection{Method: } We consider three Thinkpad W530 Laptops, one
emulating an access point (using hostapd) and two clients. We assume
client-2 requests the access point for indoor localization at $t =
6~$s. We measure the time \name\ incurs to hop between the 35 Wi-Fi
bands. Meanwhile, client-1 runs a long-lasting traffic flow. We
consider two types of flows: (1) VLC video stream over RTP; (2) TCP
flow using iperf. We repeat the experiment 30 times and find aggregate
results.

\sssection{Results: } Fig.~\ref{fig:hoptime} depicts the CDF of the
time that \name\ incurs to hop over all Wi-Fi bands. We observe that
the median hopping time is 84~ms for the Intel 5300 Wi-Fi card, in
tune with past work on other commercial Wi-Fi radios~\cite{fatvap}.

Next, Fig.~\ref{fig:viddel} plots a representative trace of the
cumulative bytes of video received over time of a VLC video stream run
by client-1 (solid blue line). {The red line plots the cumulative
  number of bytes of video played by the client. Notice that at $t =
  6~$s, there is a brief time span when no new bytes are downloaded by
  the client (owing to the localization request). However, in this
  interval, the buffer has enough bytes of video to play, ensuring
  that the user does not perceive a video stall (i.e. the blue and red
  lines do not cross). In other words, buffers in today's video
  streaming applications can largely cushion such short-lived
  outages~\cite{vidcush1, vidcush2}, minimizing impact on user
  experience}. Similarly, Fig.~\ref{fig:tcptpt} depicts a
representative trace of the throughput over time of a TCP flow at
client-1. {We observe that the TCP throughput dips only slightly by
  6.5\% at $t = 6~$s, when client-2 requests location. However, we
  emphasize that if more frequent localization is desired, we
  recommend deploying an access point or Wi-Fi beacon exclusively for
  indoor positioning. }

\subsection{Application to Personal Drones} \label{sec:droneresults}
To illustrate \name's capabilities, we evaluate how \name\ effectively
guides a personal drone to follow a user's device at an optimal
distance to take pictures.

\sssection{Method: } Our personal drone is an AscTec Hummingbird
quadrotor equipped with the AscTec Atomboard\footnote{While we use the
  atomboard due to its light-weight of only 90 grams, we note that
  \name\ is compatible with other small computing modules like the
  Intel Galileo or Fit-PC. } light-weight computing platform (with the
Intel 5300 Wi-Fi card), a Go-pro camera and a Yei-Technology motion
sensor. We 3-D print an enclosure to mount all these components safely
atop the quadrotor. Fig.~\ref{fig:droneimpl} depicts our setup. Note
that the Intel 5300 Wi-Fi card supports 3-antennas; the fourth antenna
on the quadrotor is placed only for balance and stability.

We perform our personal drone experiments in a $6~\text{m} \times
5~\text{m}$ room augmented with the VICON motion capture system
\cite{VICON}. The motion capture room uses an array of twelve infrared
cameras to track devices tagged with infrared markers at
sub-centimeter accuracy. We use the motion tracking system to find the
ground-truth trajectories of the personal drone and user device. In
each experiment, the personal drone tracks an ASUS EEPC netbook with
the Intel 5300 Wi-Fi card held by a user. The user walks along a
randomly chosen trajectory. The drone maintains a constant height and
follows the user using \name's negative-feedback loop algorithm,
described in~\xref{sec:drone} to maintain a constant distance of 1.4~m
relative to the user's device. The drone also captures photographs of
the user along the way using the Go-Pro camera mounted on the
Hummingbird quadrotor, keeping the user at 1.4~m in focus. The drone
uses the compass on the user's device and the quadrotor to ensure that
its camera always faces the user.

\sssection{Results: } Fig.~\ref{fig:dronedist} measures the CDF of
root mean squared deviation in distance of the drone relative to the
desired value of 1.4~m --- a median of 4.17~cm. {Our results reveal
  that the drone tightly maintains its relative distance to the user's
  device. Notice that our error in distance is significantly lower in
  this experiment relative to~\xref{sec:resloc}. This is because
  drones measure multiple distances as they navigate in the air, which
  helps de-noise measurements and remove outliers
  (see~\xref{sec:drone}).  }

Fig.~\ref{fig:dronetraj} depicts a candidate overhead trajectory of
the drone, captured using the Vicon motion capture system. {The
  trajectory reveals that the drone follows the user's location
  closely, as expected. Observe that at each point in its trajectory,
  the drone maintains a steady pairwise distance of 1.4~m relative to
  the device. } Finally, Fig.~\ref{fig:dronepic} depicts a
representative picture of the user that drone took along the way (face
blurred for anonymity). Notice that the picture was taken at the
optimal distance of 1.4~m and ensures that the user is at the right
focus. Note that the Go-Pro uses a wide-angle lens which ensures that
the user at 1.4~m is fully in-frame.

\section{Conclusion}
This paper presents \name, a system that measures sub-nanosecond
time-of-flight on commercial Wi-Fi radios. \name\ leverages these
time-of-flight measurements to demonstrate device-to-device indoor
positioning at state-of-the-art accuracy, without third party
support. To illustrate these capabilities, \name\ enables light-weight
personal drones to track a user's location in indoor
environments. Beyond personal drones, \name\ opens up indoor
positioning to many new contexts, where pairs of devices interact,
e.g., gesture-based gaming consoles, finding lost devices, maintaining
robotic formations, etc.

{
\let\oldthebibliography=\thebibliography
\let\endoldthebibliography=\endthebibliography
\renewenvironment{thebibliography}[1]{%
    \begin{oldthebibliography}{#1}
    \setlength{\parskip}{0ex}%
      \setlength{\itemsep}{0ex}%
    \setlength{\topsep}{10ex}
}%
{%
\end{oldthebibliography}%
}
{
\bibliographystyle{abbrv}
\bibliography{ourbib,rfid}

\begin{thebibliography}{10}

\bibitem{bosch}
Bosch laser distance measurer glm50.
\newblock
  \url{http://www.boschtools.com/Products/Tools/Pages/BoschProductDetail.aspx?pid=GLM\%2050}.

\bibitem{80211standard}
{IEEE} 802.11n-2009.
\newblock \url{http://standards.ieee.org/findstds/standard/802.11n-2009.html}.

\bibitem{VICON}
Vicon t-series.
\newblock \url{http://www.vicon.com/products/documents/Tseries.pdf}.

\bibitem{witrack}
F.~Adib, Z.~Kabelac, D.~Katabi, and R.~C. Miller.
\newblock {3D Tracking via Body Radio Reflections}.
\newblock NSDI, 2014.

\bibitem{footslam}
M.~Angermann and P.~Robertson.
\newblock {FootSLAM: Pedestrian Simultaneous Localization and Mapping Without
  Exteroceptive Sensors}.
\newblock {\em Proceedings of the IEEE}, 2012.

\bibitem{l1sparse}
F.~Bach, R.~Jenatton, J.~Mairal, and G.~Obozinski.
\newblock Convex optimization with sparsity-inducing norms, 2011.

\bibitem{ndft}
S.~Bagchi and S.~K. Mitra.
\newblock {\em The Nonuniform Discrete Fourier Transform and Its Applications
  in Signal Processing}.
\newblock Kluwer Academic Publishers, Norwell, MA, USA, 1999.

\bibitem{radar}
P.~Bahl and V.~Padmanabhan.
\newblock Radar: an in-building rf-based user location and tracking system.
\newblock In {\em INFOCOM}, 2000.

\bibitem{channelsparse}
W.~U. Bajwa, J.~Haupt, A.~Sayeed, and R.~Nowak.
\newblock Compressed channel sensing: A new approach to estimating sparse
  multipath channels.
\newblock In {\em Proceedings of the IEEE}, pages 1058--1076, 2010.

\bibitem{tof2}
T.~Bourchas, M.~Bednarek, D.~Giustiniano, and V.~Lenders.
\newblock Practical limits of wifi time-of-flight echo techniques.
\newblock In {\em IPSN}, pages 273--274, April 2014.

\bibitem{loop}
Y.~Brun, G.~Marzo~Serugendo, C.~Gacek, H.~Giese, H.~Kienle, M.~Litoiu,
  H.~M\"{u}ller, M.~Pezz\`{e}, and M.~Shaw.
\newblock {Software Engineering for Self-Adaptive Systems}.
\newblock chapter {Engineering Self-Adaptive Systems Through Feedback Loops}.
  2009.

\bibitem{ez}
K.~Chintalapudi, A.~Padmanabha~Iyer, and V.~N. Padmanabhan.
\newblock {Indoor localization without the pain}.
\newblock MobiCom, 2010.

\bibitem{chinese}
C.~Ding, D.~Pei, and A.~Salomaa.
\newblock {\em Chinese Remainder Theorem: Applications in Computing, Coding,
  Cryptography}.
\newblock World Scientific Publishing Co., Inc., River Edge, NJ, USA, 1996.

\bibitem{ndft2}
A.~Dutt and V.~Rokhlin.
\newblock Fast fourier transforms for nonequispaced data.
\newblock {\em SIAM J. Sci. Comput.}, 14(6):1368--1393, Nov. 1993.

\bibitem{fica}
J.~Fang, K.~Tan, Y.~Zhang, S.~Chen, L.~Shi, J.~Zhang, Y.~Zhang, and Z.~Tan.
\newblock Fine-grained channel access in wireless lan.
\newblock {\em IEEE/ACM Trans. Netw.}, 21(3):772--787, June 2013.

\bibitem{wifislam}
B.~Ferris~et al.
\newblock {WiFi-SLAM} using gaussian process latent variable models.
\newblock IJCAI, 2007.

\bibitem{sparsebook}
M.~Fornasier.
\newblock Numerical methods for sparse recovery.
\newblock {\em Theoretical Foundations and Numerical Methods for Sparse
  Recovery}, 14:93--200, 2010.

\bibitem{pulse}
S.~Gezici, Z.~Tian, G.~B. Biannakis, H.~Kobayashi, A.~F. Molisch, V.~Poor,
  Z.~Sahinoglu, S.~Gezici, Z.~Tian, G.~B. Giannakis, H.~Kobayashi, A.~F.
  Molisch, H.~V. Poor, and Z.~Sahinoglu.
\newblock Localization via ultra-wideband radios.
\newblock In {\em IEEE Signal Processing Magazine}, pages 70--84, 2005.

\bibitem{caesar}
D.~Giustiniano and S.~Mangold.
\newblock Caesar: Carrier sense-based ranging in off-the-shelf 802.11 wireless
  lan.
\newblock CoNEXT, 2011.

\bibitem{phaser}
J.~Gjengset, J.~Xiong, G.~McPhillips, and K.~Jamieson.
\newblock Phaser: Enabling phased array signal processing on commodity wifi
  access points.
\newblock MobiCom, 2014.

\bibitem{ndft1}
L.~Greengard and J.~yub Lee.
\newblock Accelerating the nonuniform fast fourier transform.
\newblock {\em SIAM REVIEW}, 46(3):443--454, 2004.

\bibitem{reciprocity}
M.~Guillaud, D.~Slock, and R.~Knopp.
\newblock A practical method for wireless channel reciprocity exploitation
  through relative calibration.
\newblock {\em ISSPA}, 2005.

\bibitem{csitool}
D.~Halperin, W.~Hu, A.~Sheth, and D.~Wetherall.
\newblock Tool release: Gathering 802.11n traces with channel state
  information.
\newblock {\em ACM SIGCOMM CCR}, 2011.

\bibitem{chinesefourier}
M.~Hazewinkel.
\newblock {\em Encyclopaedia of Mathematics: An Updated and Annotated
  Translation of the Soviet "Mathematical Encyclopaedia}.
\newblock Encyclopaedia of Mathematics. Springer Netherlands, 1997.

\bibitem{ofdmbook}
J.~Heiskala and J.~Terry, Ph.D.
\newblock {\em OFDM Wireless LANs: A Theoretical and Practical Guide}.
\newblock Sams, Indianapolis, IN, USA, 2001.

\bibitem{proxconv}
K.~Hou, Z.~Zhou, A.~M.-C. So, and Z.-Q. Luo.
\newblock On the linear convergence of the proximal gradient method for trace
  norm regularization.
\newblock In {\em Advances in Neural Information Processing Systems 26}, pages
  710--718. Curran Associates, Inc., 2013.

\bibitem{vidcush2}
T.-Y. Huang, R.~Johari, and N.~McKeown.
\newblock {Downton Abbey Without the Hiccups: Buffer-based Rate Adaptation for
  HTTP Video Streaming}.
\newblock FhMN, 2013.

\bibitem{vidcush1}
T.-Y. Huang, R.~Johari, N.~McKeown, M.~Trunnell, and M.~Watson.
\newblock {A Buffer-based Approach to Rate Adaptation: Evidence from a Large
  Video Streaming Service}.
\newblock SIGCOMM, 2014.

\bibitem{chancont}
A.~T. Islam and I.~Misra.
\newblock Performance of wireless ofdm system with ls-interpolation-based
  channel estimation in multi-path fading channel.
\newblock IJCSA, 2012.

\bibitem{PinPoint}
K.~Joshi, S.~Hong, and S.~Katti.
\newblock Pinpoint: Localizing interfering radios.
\newblock NSDI, 2013.

\bibitem{fatvap}
S.~Kandula, K.~C.-J. Lin, T.~Badirkhanli, and D.~Katabi.
\newblock {FatVAP}: Aggregating {AP} backhaul capacity to maximize throughput.
\newblock In {\em NSDI}, 2008.

\bibitem{harmonica}
B.~Kempke, P.~Pannuto, and P.~Dutta.
\newblock {Harmonia: Wideband Spreading for Accurate Indoor RF Localization}.
\newblock HotWireless, 2014.

\bibitem{tracksense}
M.~Kohler, S.~N. Patel, J.~W. Summet, E.~P. Stuntebeck, and G.~D. Abowd.
\newblock Tracksense: Infrastructure free precise indoor positioning using
  projected patterns.
\newblock PERVASIVE, 2007.

\bibitem{ubicarse}
S.~Kumar, S.~Gil, D.~Katabi, and D.~Rus.
\newblock Accurate indoor localization with zero start-up cost.
\newblock MobiCom, 2014.

\bibitem{tof4}
S.~Lanzisera, D.~Zats, and K.~Pister.
\newblock Radio frequency time-of-flight distance measurement for low-cost
  wireless sensor localization.
\newblock {\em Sensors Journal, IEEE}, 11(3):837--845, March 2011.

\bibitem{lietal}
F.~Li, C.~Zhao, G.~Ding, J.~Gong, C.~Liu, and F.~Zhao.
\newblock {A Reliable and Accurate Indoor Localization Method Using Phone
  Inertial Sensors}.
\newblock UbiComp, 2012.

\bibitem{guoguo}
K.~Liu~et al.
\newblock Guoguo: Enabling fine-grained indoor localization via smartphone.
\newblock MobiSys, 2013.

\bibitem{conexttof}
A.~Marcaletti, M.~Rea, D.~Giustiniano, V.~Lenders, and A.~Fakhreddine.
\newblock Filtering noisy 802.11 time-of-flight ranging measurements.
\newblock CoNEXT, 2014.

\bibitem{sail}
A.~T. Mariakakis, S.~Sen, J.~Lee, and K.-H. Kim.
\newblock Sail: Single access point-based indoor localization.
\newblock MobiSys, 2014.

\bibitem{tof5}
K.~Muthukrishnan, G.~Koprinkov, N.~Meratnia, and M.~Lijding.
\newblock Using time-of-flight for wlan localization: feasibility study, 2006.

\bibitem{markers}
Y.~Nakazato, M.~Kanbara, and N.~Yokoya.
\newblock Localization system for large indoor environments using invisible
  markers.
\newblock VRST, 2008.

\bibitem{singledevicefree}
A.~Popleteev.
\newblock Device-free indoor localization using ambient radio signals.
\newblock UbiComp '13 Adjunct, 2013.

\bibitem{personal}
B.~Popper.
\newblock The drone you should buy right now.
\newblock
  \url{http://www.theverge.com/2014/7/31/5954891/best-drone-you-can-buy}.

\bibitem{wisee}
Q.~Pu, S.~Gupta, S.~Gollakota, and S.~Patel.
\newblock Whole-home gesture recognition using wireless signals.
\newblock MobiCom, 2013.

\bibitem{sourcesync}
H.~Rahul, H.~Hassanieh, and D.~Katabi.
\newblock {SourceSync: A Distributed Wireless Architecture for Exploiting
  Sender Diversity}.
\newblock In {\em ACM SIGCOMM 2010}, New Delhi, India, August 2010.

\bibitem{megamimo}
H.~Rahul, S.~Kumar, and D.~Katabi.
\newblock {MegaMIMO: Scaling Wireless Capacity with User Demands}.
\newblock In {\em ACM SIGCOMM 2012}, Helsinki, Finland, August 2012.

\bibitem{zee}
A.~Rai, K.~K. Chintalapudi, V.~N. Padmanabhan, and R.~Sen.
\newblock Zee: zero-effort crowdsourcing for indoor localization.
\newblock Mobicom '12.

\bibitem{theorytoa}
S.~Ravindra and S.~N. Jagadeesha.
\newblock Time of arrival based localization in wireless sensor networks : {A}
  linear approach.
\newblock {\em CoRR}, abs/1403.6697, 2014.

\bibitem{tof3}
L.~Schauer, F.~Dorfmeister, and M.~Maier.
\newblock Potentials and limitations of wifi-positioning using time-of-flight.
\newblock In {\em IPIN}, 2013.

\bibitem{adapt}
S.~Shin, C.~Park, J.~Kim, H.~Hong, and J.~Lee.
\newblock Adaptive step length estimation algorithm using low-cost mems
  inertial sensors.
\newblock In {\em SAS}, 2007.

\bibitem{cricket}
A.~Smith~et al.
\newblock {Tracking Moving Devices with the Cricket Location System}.
\newblock In {\em MobiSys}, 2004.

\bibitem{Tse}
D.~Tse and P.~Vishwanath.
\newblock {\em Fundamentals of Wireless Communications}.
\newblock Cambridge University Press, 2005.

\bibitem{aoaest}
F.~Wen and C.~Liang.
\newblock {An Indoor AOA Estimation Algorithm for IEEE 802.11ac Wi-Fi Signal
  Using Single Access Point}.
\newblock {\em Communications Letters, IEEE}, 2014.

\bibitem{sap1}
F.~Wen and C.~liang.
\newblock Fine-grained indoor localization using single access point with
  multiple antennas.
\newblock {\em Sensors Journal, IEEE}, 2015.

\bibitem{tof1}
S.~B. Wibowo, M.~Klepal, and D.~Pesch.
\newblock Time of flight ranging using off-the-self ieee802.11 wifi tags.
\newblock POCA, 2009.

\bibitem{arraytrack}
J.~Xiong and K.~Jamieson.
\newblock Arraytrack: A fine-grained indoor location system.
\newblock NSDI '13, 2013.

\bibitem{synchronicity}
J.~Xiong, K.~Jamieson, and K.~Sundaresan.
\newblock Synchronicity: Pushing the envelope of fine-grained localization with
  distributed mimo.
\newblock HotWireless, 2014.

\bibitem{horus}
M.~Youssef and A.~Agrawala.
\newblock The {Horus} {WLAN} location determination system.
\newblock MobiSys, 2005.

\end{thebibliography}
}
}

\end{sloppypar}
\end{document}